\documentclass{amsart}
	   \usepackage[pdftex]{graphicx}
\usepackage{amsfonts}
\usepackage{amsaddr}
\usepackage{algorithm}
\usepackage{algorithmicx}
\usepackage{algpseudocode}
\usepackage{paralist}
\usepackage{array}
\usepackage{url}
\usepackage{siunitx}
	\usepackage{calc}
\usepackage{footmisc}
\begin{document}
\begin{abstract}
In mobile ad-hoc networks, neighbor discovery protocols are used to find surrounding devices and to establish a first contact between them. Since the clocks of the devices are not synchronized and their energy-budgets are limited, usually duty-cycled, asynchronous discovery protocols are applied.
Only if two devices are awake at the same point in time, they can rendezvous. Currently, time-slotted protocols, which subdivide time into multiple intervals with equal lengths (slots), are considered to be the most efficient discovery schemes. 
In this paper, we break away from the assumption of slotted time. We propose a novel, continuous-time discovery protocol, which temporally decouples beaconing and listening. Time is continuous, which means that each device periodically sends packets with a certain interval that can be chosen freely in arbitrarily small steps. These points in time are independent from the time instances the device listens to the channel. Similarly, each device has a listening interval with which it repeatedly switches on its receiver for a certain amount of time. Unlike in slotted protocols, both interval lengths, their temporal offsets and the listening-duration in each interval are independent from each other.
By optimizing these interval lengths, we show that this scheme can, to the best of our knowledge, outperform all known slotted protocols such as DISCO, U-Connect or Searchlight significantly. For example, Searchlight takes up to $\SI{1020}{\percent}$ longer than our proposed technique to discover a device with the same duty-cycle and hence energy-consumption. Further, our proposed technique can also be applied in widely-used asymmetric purely interval-based protocols such as ANT or Bluetooth Low Energy, thereby optimizing their energy-consumptions.
\end{abstract}

\title{Slotless Protocols for Fast and Energy-Efficient Neighbor Discovery}
\author{Philipp H. Kindt, Marco Saur and Samarjit Chakraborty} 
\address[]{Institute for Real-Time Computer Systems,\\ Technische Universit\"at M\"unchen,\\ 80290, Munich, Germany.}
\email{kindt\_OR\_saur\_OR\_chakraborty@rcs.ei.tum.de}

\maketitle

\section{Introduction}
\label{sec:introduction}
Low power mobile ad-hoc networks (MANETs), which provide wireless connectivity between multiple mobile devices without the need for stationary, grid-powered infrastructure, are widely used. Since the participating devices are battery-powered, energy-saving communication is a crucial requirement, which is usually achieved through duty-cycled protocols. Such protocols allow the hardware to go to a sleep mode during most of the time, and to wake up only if packets need to be exchanged. Once a connection between two devices has been established, a common wakeup schedule applies, which is known to both sides. Therefore, both devices always wakeup simultaneously and sleep in the meantime. However, before initiating a connection, the clocks of the devices are not synchronized and a common wakeup schedule cannot be realized. For establishing a first contact, periods of time at which only one device is active whereas the other one is asleep cannot be avoided, if duty-cycling is used. To minimize these periods and to ensure a reasonably fast discovery procedure, multiple neighbor discovery protocols have been proposed. They define optimized wakeup schedules for both devices, thereby minimizing the worst-case discovery latency and at the same time reducing the duty-cycle and hence the energy-consumption. 

\begin{figure}[t]
	\centering
	\includegraphics[width=\linewidth]{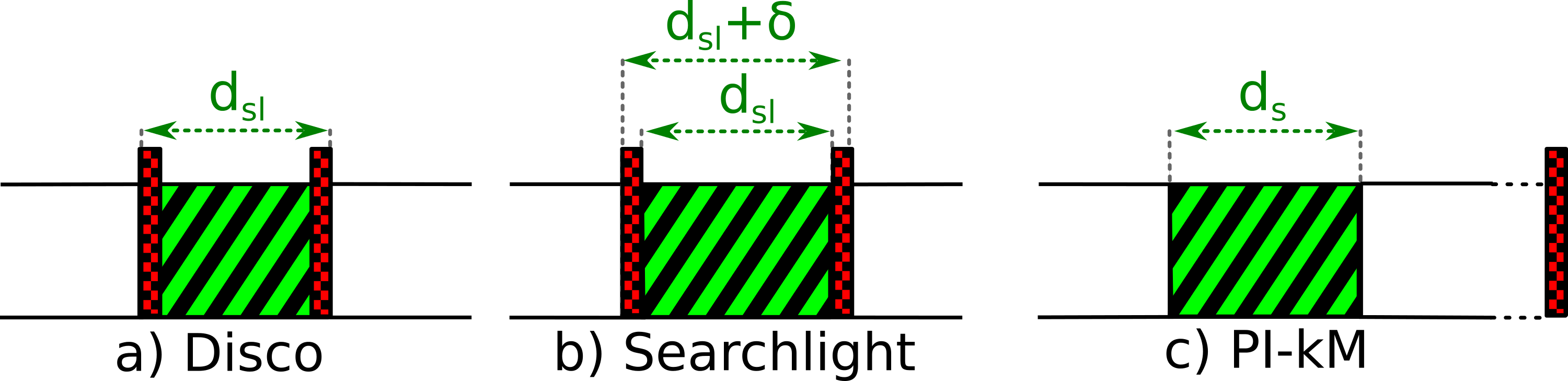}
	\caption{Slot design of a) Disco \protect\cite{dutta:08}, b) Searchlight \protect\cite{bakht:12} and $\mathbf{PI-kM}$.}
	\label{fig:disco_searchlight_pi_kmp_slot_strategy} 
\end{figure}

The large majority of the recently proposed protocols assume time to be slotted: A certain period $T$ is subdivided into a number of $i$ intervals with equal lengths $d_{sl}$. In each of these slots, a device can either be in a sleep-mode or active. In each active slot, a device transmits a beacon at the beginning and the end of the slot and listens for incoming beacons in between of them, as shown in Figure~\ref{fig:disco_searchlight_pi_kmp_slot_strategy} a). Accordingly, whenever two active slots of a pair of devices overlap temporally for at least the duration of one beacon $d_a$, the beacons are received successfully at both devices and the discovery is complete.

\begin{figure}[b]
\centering
\includegraphics[width=\linewidth]{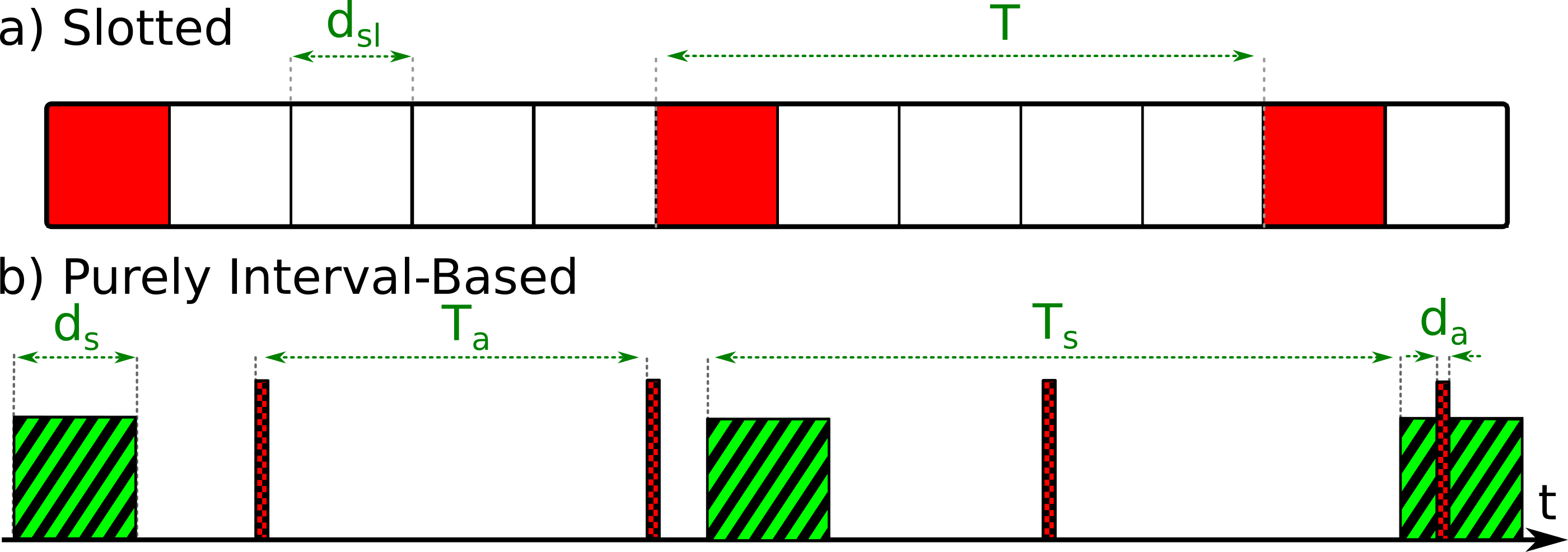}
\caption{Slotted (a) and purely interval-based (b) discovery.}
\label{fig:interval_vs_slotted} 
\end{figure}
Figure~\ref{fig:interval_vs_slotted} a) depicts the schedule of such a slotted protocol. Active slots are colored and passive slots are left white. The objective of neighbor discovery protocols is to determine a pattern of active slots for which a deterministic (i.e., discovery is guaranteed within a certain number of slots) maximum latency is always achieved for every possible time-offset between two devices following such a schedule. Multiple deterministic schedules have been proposed, which are based on coprime interval lengths \cite{dutta:08}, quorum-patterns \cite{Tseng:02}, systematic probing \cite{bakht:12} or optimal difference codes \cite{meng:14}. We describe these techniques in more detail in Section~\ref{sec:related_work}.

An alternative to time-slotted protocols is contiuous-time, periodic-interval (PI)-based discovery, as shown in Figure~\ref{fig:interval_vs_slotted} b). Unlike the slotted protocols mentioned above, all known implementations are asymmetric, in the sense of assigning different roles to different devices. One device, which is called the \textit{advertiser}, periodically sends out packets with a period of $T_a$ time-units. The transmission duration of a packet $d_a$ is determined by the number of bytes sent. The other device is referred to as the \textit{scanner}. It periodically switches on its receiver for a duration called the \textit{scan window} $d_s$. The repetition period is called the \textit{scan interval} $T_s$. Time is assumed to be continuous, beaconing and receiving occur independently from each other and no slots exist. Such protocols are also widely used in practice. They have been proposed first in \cite{schurgers:02}, with separate channels for beaconing and for responses to the received packets. The unidirectional mode of the ANT/ANT+\cite{AntSpec:14} protocol makes use of the scheme described without any modifications. The most popular protocol that relies on this technique is Bluetooth Low Energy (BLE), which applies two slight modifications to it. First, beacons are sent on up to three different channels in a row after each advertising interval $T_a$. The scanner cycles between these three channels after each scan-interval. Second, a random delay $\rho \in [\SI{0}{s}, \SI{10}{ms}]$ is added to $T_a$ in each period.

Whereas much research has been carried out to improve time-slotted protocols, significantly less work has been presented on the optimization of PI-based ones. The main reason is that until recently, the popular belief was that they cannot guarantee deterministic discovery. Attempts to find special configurations for BLE, e.g., coprime interval lengths \cite{kandhalu:13} which provide upper latency bounds, did not turn out to be a promising alternative to time-slotted protocols. Except for some special-cases, such protocols have not been well understood, and neither the mean discovery latencies nor any information on upper bounds could be derived with analytical methods. Recently, in \cite{kindt:15}, the first comprehensive analysis of purely interval-based protocols has been proposed. It revealed that such protocols provide upper latency bounds for all interval lengths except a finite number of singularities. Therefore, they can be applied in applications in which slotted protocols are used nowadays to ensure deterministic discovery. However, their maximum performances, especially for optimized parametrizations, are still not clear.

The main contribution of this paper is proposing an optimization technique for PI-based protocols, which can be used for computing optimized parameter values for $T_a$, $T_s$ and $d_s$. The resulting latencies for a given duty-cycle are very short. To the best of our knowledge, for reasonable slot lengths of a few milliseconds, the $PI-kM$-protocol can outperform all known time-slotted protocols significantly. For instance, for certain choices of duty cycles, Searchlight takes up to $10.2 \times$ longer for discovering a device with the same duty-cycle and hence the same energy-consumption. For no choice of duty-cycles, Searchlight outperforms ours. Further, as already mentioned, our proposed theory can be used to optimize the parametrizations of existing PI-based protocols such as ANT/ANT+ and, to some extend, also BLE. In addition, unlike slotted protocols, our proposed protocol can realize any specified duty-cycle. 
The main insight we have exploited is temporally decoupling beaconing from listening. This allows a) for sending additional beacons compared to slotted protocols and b) for sending the beacons at optimal points in time.

PI-based protocols have three degrees of freedom ($T_a$ , $T_s$ and $d_s$), and the maximum latency is a very irregular function with a large number of minima and maxmia, as can be seen in Figure~\ref{fig:PI0MTaValues}. Evaluating all possible valuations exhaustively to find optimal parametrizations cannot be realized in reasonable amounts of time. Further, the models from the literature~\cite{kindt:15} for computing the upper latency bounds are in the form of recursive algorithms, which makes it impossible to apply differential methods. As a result, finding optimal parametrizations is a difficult task which has not yet been solved.

For computing the discovery-latency of PI-based protocols, the shrinkage of the temporal offset between appropriate neighboring advertising packets and scan windows is tracked over time. 
The theory in \cite{kindt:15} reveals that there are multiple orders of latency-maxima and minima, as described in Section \ref{sec:related_work} in detail. In brief, the order of a maximum/minimum refers to the number of advertising- and scan intervals that lie in between a neighboring pair of an advertising packet and a scan window. For example, if $T_a < T_s$ and if only one instance of $T_a$ and $T_s$ are considered, the distance between each advertising packet and the next temporally right scan-event shrinks in multiples of $T_a$ for subsequent advertising packets (cf. Figure \ref{fig:simpleLatencyComputationModel} a)). This is referred to as an order-0 process. Whereas for some initial offsets the order-0 process might lead to a match after a certain number of $T_a$ - intervals, for other offsets, linear combinations of the same pair of intervals need to be accounted for. The next higher neighborhood-relationship that needs to be examined is $T_s$ and multiples of $\lfloor \frac{T_s}{T_a} \rfloor T_a$ (or $\lceil \frac{T_s}{T_a} \rceil T_a$, depending on the situation). Such a situation is depicted in Figure \ref{fig:modeling_pi}, in which the temporal offset $\Phi[k]$ shrinks for increasing numbers of scan-intervals $k$. This is referred to as an order-1 process. Processes of higher order exist, which require computations with more elaborate linear combinations.
 
In this paper, we propose an optimization technique to compute parameter values which always lie in latency minima of orders 0 and 1. These minima provide pareto-optimal discovery latencies (under the constraint that only maxima of order 0 and 1 are considered). The parameter optimization works as follows: First, we derive a linear function of $T_s$ to compute all advertising intervals that lead to latency minima of order 0, as described in detail in Section \ref{sec:PIkMProtocol}. Based on this function, we deduce optimal values of $T_s$. After these steps, the interval lengths $T_a$ and $T_s$ are parametrized by two integer numbers $k$ and $M$. Based on the desired target duty-cycle, $k$, $M$ and the scan window $d_s$ are optimized using differential methods. We call solutions using these optimizations \textit{$PI-kM$ - protocols} and use them for the comparison against recent slotted protocols, such as DISCO \cite{dutta:08}, U-connect \cite{Kandhalu:10} and Searchlight \cite{bakht:12}. In this comparison, we also take into account optimal difference codes \cite{meng:14}, even though they are currently not realizable except for some special configurations. In addition, we compare our proposed protocols against Lightning \cite{wei:16}, even though no real-world implementation of Lightning has been presented, yet. Finally, the recently presented protocol Nihao \cite{qiu:16} is considered, since it applies a theory of pseudo slots, which can be seen as an intermediate step towards fully slotless protocols, as considered in this paper. As already mentioned, our proposed protocol can outperform all of these protocols in terms of discovery-latency for any given duty-cycle.

Comparing the performance of slotless, PI-based protocols against slotted ones, as carried out in this paper, is challenging. Mainly, this is because of the following reasons.
\begin{enumerate}
	
\item PI- based and slotted protocols are not directly comparable, because all known purely interval-based protocols assign different roles to each device (viz., advertiser and scanner), whereas slotted protocols typically assume that every device implements both roles. To achieve comparability, we propose the first symmetric, two-way variant of such PI-protocols, in which both devices perform both advertising and scanning. Known asymmetric protocols such as ANT/ANT+ can be described as special cases of this protocol by setting some of its parameters to infinity. Therefore, all our results can directly be applied to them and we, for the first time, present a mathematical framework for efficient parametrizations of these protocols.

\item Third, PI-based protocols guarantee a maximum latency in terms of time-units, whereas slotted solutions achieve deterministic discovery within a given number of slots. Clearly, their discovery latency depends on the slot size $d_{sl}$. Towards a fair comparison, we determine a lower bound for reasonable slot lengths.
\end{enumerate}

The rest of this paper is organized as follows. In the next section, we provide an overview on the state of the art of slotted and PI-based discovery protocols. Next, in Section \ref{sec:PIkMProtocol}, we describe our proposed protocol and its optimizations in detail, including a theoretical model on the maximum discovery latency. We evaluate our theory by measurements on a real-world implementation of the $PI-kM$-protocol and present a performance-comparison with time-slotted protocols in Section \ref{sec:evaluation}. Finally, we discuss the implications of our theory in Section \ref{sec:conclusion}.

\section{Related Work}
\label{sec:related_work}

In this section, we first give a brief overview on time-slotted discovery protocols. Next, we present related work on purely interval-based discover and compare the relevant related work with this paper.
\subsection{Time-Slotted Discovery}
Within the last decade, a large number of slotted discovery protocols have been proposed. As already mentioned, they subdivide a certain time period $T$ into multiple equal-length slots. Each protocol defines a unique discovery schedule, which determines the set of active slots in each period. Since this paper focuses on \textit{symmetric asynchronous neighbor discovery}, i.e. neighbor discovery in which all nodes follow the same schedule with independent clocks, we only present related work on symmetric neighbor discovery.

\begin{figure}[t]
\centering
\includegraphics[width=\linewidth]{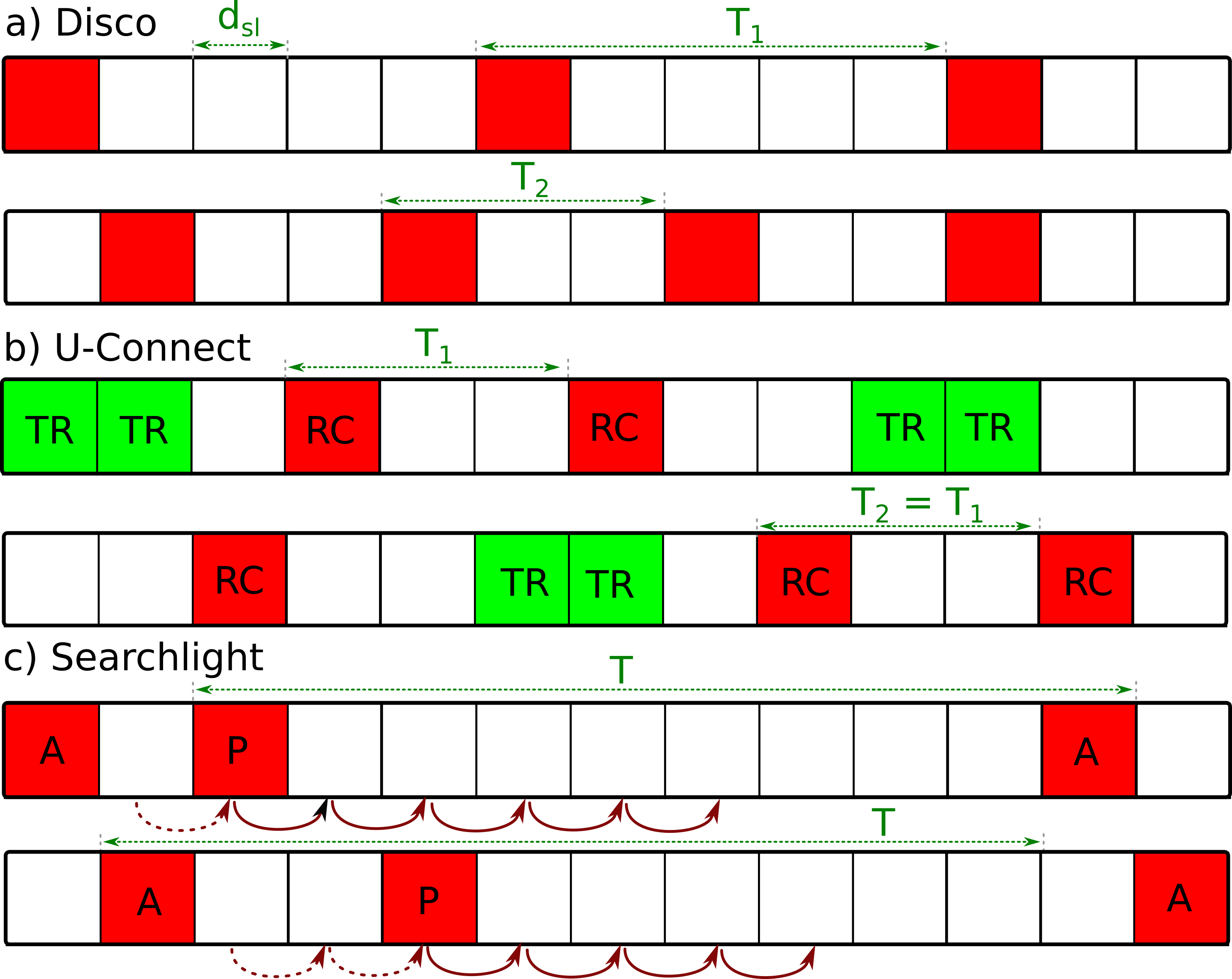}
\caption{Discovery Schedules of a) Disco \protect\cite{dutta:08}, b) U-Connect \protect\cite{Kandhalu:10} and c) Searchlight \protect\cite{bakht:12}.}
\label{fig:disco_searchlight_uconnect_scheme} 
\end{figure}

One of the first slotted protocols has been proposed in \cite{Tseng:02}, which is based on a quorum schedule to achieve determinism. 
A different approach which achieves similar latencies is Disco \cite{dutta:08}. Its schedule for two devices that attempt to discover each other is shown in Figure \ref{fig:disco_searchlight_uconnect_scheme} a). Disco assumes that each device chooses two periods $T_1, T_2$ (e.g., $T_1 = 5$ and $T_2 = 3$ in Figure \ref{fig:disco_searchlight_uconnect_scheme}), which consist of a prime number of slots. As can be seen in the figure, such a schedule overlaps guaranteed after a limited amount of time. The Chinese Remainder Theorem states that the worst-case latency is the product of the coprime periods of both devices. 

However, two devices which have chosen the same pair of prime numbers risk never finding each other. This problem can be overcome by U-Connect \cite{Kandhalu:10}, which is depicted in Figure \ref{fig:disco_searchlight_uconnect_scheme} b). Each device choses a coprime interval length of $T_1$ or $T_2$ slots, respectively. In each period, the first slot is active. In addition, each device has a super-period of $T_1^2$ or $T_2^2$ slots. The first $\left\lfloor \frac{T + 1}{2} \right\rfloor$ slots of each super-period are also active. As can be seen easily, also in cases with $T_1$ = $T_2$, mutual discovery is guaranteed. In symmetric settings (i.e., both devices have the same duty-cycle), U-Connect provides lower latency-duty-cycle products than Disco, which means that for a given duty cycle, the discovery can be carried out in less time.
Unlike Disco (cf. Figure \ref{fig:disco_searchlight_pi_kmp_slot_strategy} a)), the implementation of U-Connect presented in \cite{Kandhalu:10} uses two different types of slots. During the regular slots of the interval $T$, the device is in reception mode (indicated by \textit{RC} in Figure \ref{fig:disco_searchlight_uconnect_scheme}~b), whereas the device continuously transmits within the active slots of the hyper-period $T^2$ (indicated by \textit{TR} in the Figure).  
Another significant improvement of the latency-duty-cycle product has been achieved by Searchlight \cite{bakht:12}. Like in Disco, each active slot consists of a listening period, which is preceded and terminated by sending a beacon packet. However, as shown in Figure \ref{fig:disco_searchlight_pi_kmp_slot_strategy}~b), active slots are by a percentage $\delta$ longer than passive slots (so-called \textit{over-length slots}). The schedule of Searchlight is shown in Figure \ref{fig:disco_searchlight_uconnect_scheme}~c). Again, there is a period $T$ and a hyper-period $T^2$. In this paper, we consider the symmetric case in which both devices have the same period. The first slot of each period is referred to as an \textit{anchor slot} A. In addition, there is another active slot per period, the so-called \textit{probe slot} P. In the first period, the second slot is a probe-slot. In each succeeding period, the position of the probe-slot is increased by one until reaching half of the interval $T$. With this scheme, an upper latency-bound of approximately $\frac{T^2}{2}$ slots is achieved. With the over-length slots shown in Figure \ref{fig:disco_searchlight_pi_kmp_slot_strategy}~b), every second probe-slot can be skipped and therefore a maximum latency of coarsely $\frac{T^2}{4}$ slots is guaranteed. This version of Searchlight is usually referred to as \textit{Searchlight-Striped} or \textit{Searchlight-S}.

Further work has been carried out recently to reduce the latencies of slotted discovery protocols, of which we briefly present a selection below. In \cite{Sun:14}, a framework called HELLO to construct slotted discovery protocols has been proposed. Disco, U-Connect and Searchlight can be constructed as special cases of HELLO. According to \cite{Sun:14}, Searchlight achieves the lowest discovery-latencies for symmetric duty-cycles. \cite{zhang:12} proposes to apply additional slots in protocols like Disco to exchange informations on already known neighbors. In \cite{chen:15}, two new slotted protocols called HEDIS and TODIS have been proposed, which outperform Searchlight in asymmetric cases. For symmetric discovery, Searchlight still performs better. Further, \cite{meng:14} combines overflowing slot lengths (as described for Searchlight) with optimal difference codes. Whereas optimal codes can only be realized for a few target duty-cycles, an algorithm to create approximations (with reduced performance) for every target-duty-cycle has been presented, which provides better latency-duty-cycle products than Searchlight. 
In \cite{chen:15}, a scheme to construct schedules which exploit over-length slots in any given schedule for non-over-length slots such as Disco or U-Connect has been proposed. Another recently proposed discovery protocol, which claims to achieve shorter latencies than Searchlight, is Lightning \cite{wei:16}. However, the results rely entirely on simulation and therefore, it is not clear how it performs in implementations, yet. To the best of our knowledge, together with optimal difference codes, Lightning provides the lowest latency bounds of all currently known protocols.

A work that has been presented recently is Nihao \cite{qiu:16}. Unlike the previously known slotted protocols, it defines dedicated receive and transmit slots. In each receive slot, the radio listens to the channel during the whole slot length. In each transmit slot, one beacon is sent at the beginning of the slot, and the device goes back to the sleep mode afterwards. This can be seen as a \textit{pseudo-slot}, which leads a to significantly different protocol behavior than known from slotted solutions. Therefore, Nihao can be seen as an intermediate step between slotted and fully slotless protocols, as studied in this paper. Hence, we provide a more detailed comparison in what follows.

Three different versions of Nihao have been proposed. For the first one, \textit{S-Nihao}, the authors state that it requires the fraction of the packet transmission duration and the slot length to be smaller than its duty-cycle. Therefore, it is restricted to relatively large duty-cycles or to very long slot lengths \cite{qiu:16}, which is infeasible in practice.

The second version \textit{G-Nihao} offers two parameters $m$ and $n$, which can be used to adjust the number of beacons and listening slots of each period. While the theory of pseudo-slots used in G-Nihao and the fully slotless theory in this paper results in similar schedules as one of the three proposed protocol variants in this paper (e.g., the PI-0M-protocol), there are two main differences:
\begin{compactenum}
	
\item G-Nihao does not contain any parameter-optimizations of $m$ and $n$ for maximum performance. Therefore, it is not clear how to chose them to achieve the shortest discovery latency at a given duty-cycle. It has been shown that pseudo-slotted protocols can outperform existing slotted protocols when assuming the same channel utilization, and that the performance can be increased with additional numbers of beacons sent \cite{qiu:16}. However, the optimal number of beacons for reaching the best discovery-latencies and the corresponding performance is not clear. In contrast, our proposed theory contains built-in optimizations of all parameters, therefore minimizing the latency for a given duty-cycle. The discovery-latencies achieved are significantly lower than those of all known protocols, including G-Nihao for the parametrizations presented in \cite{qiu:16}. As we will show, the resulting channel-occupancy is increased compared to existing protocols, but always below a reasonable bound of $\approx \SI{4}{\percent}$. 
\item Because schedules constructed by the theory presented in \cite{qiu:16} are based on slotted time, they can only realize certain duty-cycles. As we will show, this is a limitation in practice. In contrast, with our proposed theory, practically any specified duty-cycle can be realized.
\end{compactenum}
The third version of Nihao, B-Nihao, is a special case of G-Nihao, and therefore shares its major properties.

\subsection{Purely Interval-Based Discovery}
As already mentioned, PI-based protocols are also widely used in practice, but significantly less research has been carried out on them until recently. The main reason is that their analysis is significantly more complex than for slotted ones. Therefore, only Monte-Carlo simulations were available for over a decade. With Bluetooth Low Energy applying a PI-based scheme, its discovery latency became of high interest and multiple simulation models and -results have been presented \cite{kindt:13}, \cite{mokhtari:15}. These simulation results have revealed that for varying parameter values, the mean latency is a complex curve with a large number of maxima and minima. Therefore, \cite{kindt:13} concludes that modeling and optimizing these parameters is a crucial requirement for low-power discovery, since it makes a large difference if parameter values lead to a peak or lie within a minimum. Further, a model for the special case of $T_a < d_s$ has been proposed in \cite{schurgers:02} and adapted to Bluetooth Low Energy in \cite{liu:12_short}, \cite{liu:12_long}, \cite{liu:12_techrep}. Since discrete-event simulations are not suitable for determining upper latency bounds, the popular belief was that PI-based protocols cannot guarantee any upper latency bounds in general. Therefore, special parametrizations which fulfill the Chines Remainder Theorem for BLE have been proposed \cite{kandhalu:13}. 
\begin{figure}[tbh]
\centering
\includegraphics[width=\linewidth]{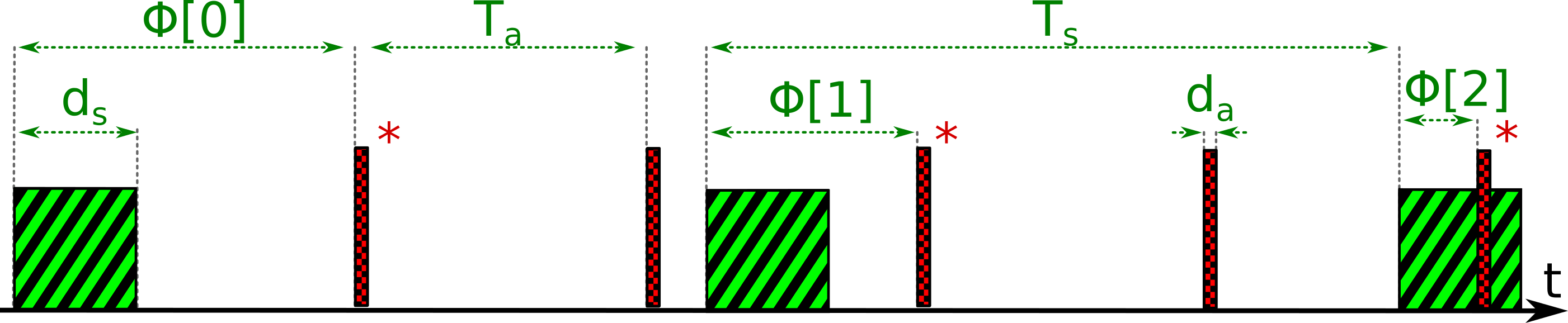}
\caption{Modeling PI-based protocols \protect\cite{kindt:15}.}
\label{fig:modeling_pi} 
\end{figure}

Recently, a novel modeling technique has been presented, which is capable of describing the mean- and worst-case discovery latencies of PI-based protocols \cite{kindt:15}. Our paper is built upon this theory. The main idea of such a model is shown in Figure \ref{fig:modeling_pi}. The ruled squares show the scan windows, whereas the hatched vertical bars depict the advertising packets. When the advertiser comes into range of the scanner, the first advertising packet has a temporal offset of $\Phi[0]$ to the first scan-event considered. The subsequent packets are either classified as closest right neighbors (marked with a * in the Figure), or as remaining ones. In this example, only the closest right neighbors have a chance on a successful reception. The temporal distance $\Phi[i]$ between closest right neighbors and their (left) neighboring scan window shrinks for increasing values of $i$, where the index $i$ identifies the instance of the larger interval out of $(T_s, T_a)$. The amount of shrinkage per interval is constant, such that it is:
\begin{equation}
\Phi[i + 1] - \Phi[i] = \gamma.
\end{equation} 
A successful reception occurs once $\Phi[i]$ has shrunken below the length of one scan-window $d_s$. 
For the worst case latency, all possible values of the initial offset $\Phi[0]$ need to be taken into account. For the intervals exemplified in Figure \ref{fig:modeling_pi}, the maximum latency is
\begin{equation}
\label{eq:discoveryLatencyWCSh}
d_{m} = \underbrace{\left\lceil \frac{T_s - T_a}{T_a} \right\rceil T_a}_{A} + \underbrace{\left\lceil \frac{T_a - d_s + d_a}{\gamma} \right\rceil \cdot \left\lceil \frac{T_s}{T_a} \right\rceil  T_a}_{B} + d_a,
\end{equation}
with 
\begin{equation}
\label{eq:gammaSh}
\gamma = \left\lceil \frac{T_s}{T_a} \right\rceil \cdot T_a - T_s.
\end{equation}
Term A accounts for cases in which $\Phi[0] > T_a$. In such cases, the distance $\Phi$ is increased in multiples of $T_a$ time-units, until an advertising packet which is temporally on the right of the second scan window is reached. From there on, $\Phi[1]$ is measured against the second scan window and becomes smaller than $T_a$. The remaining distance $\Phi[1] \leq T_a$ is shrunken in multiples of $\gamma$ until the next scan-window is reached. The latency caused by this shrinkage is accounted for by term B.
However, there are cases in which the amount of shrinkage $\gamma$ exceeds $d_s$ and accordingly, the advertising packet might temporally "overtake" the scan window. In such cases, appropriate linear combinations of both intervals need to be considered to compute another, effective parameter $\gamma_2 < \gamma$. This is called a \textit{higher-order $\mathit{\gamma}$-process}. The example of Figure~\ref{fig:modeling_pi} is a order-1 process, which implies that $\gamma < d_s$ and therefore, "overtaking" the scan window cannot occur. Similarly, we refer to order-0 processes as the growth of $\Phi[i]$ in multiples of $T_a$.  Modeling higher-order processes is more elaborate and we refer to \cite{kindt:15} for more details. For this paper, only order-0 and order-1-processes are relevant.  Further, there are situations in which the distance $\Phi[i]$ grows in multiples of $\gamma$ with increasing numbers of $i$. They can be handled similar to shrinking situations by tracking the distance to the next (succeeding) scan window.

Although the theory presented in \cite{kindt:15} provides a closed-form formulation of the discovery latency, it cannot be used for parameter optimizations, since its functions are not well-behaved. One of the main technical contributions of this paper is to develop an equivalent formulation of this theory. It covers the relevant cases, only, but is amenable to systematical parameter optimizations. Based on this simplified model, we 
derive optimized parametrizations, as described in Section \ref{sec:PIkMProtocol}.

\textbf{Our Contributions:} Compared to the literature, we make the following contributions:
\begin{itemize}
\item For comparing PI-based protocol against time-slotted ones, we propose a novel symmetric discovery protocol, which is based on continuous-time, periodic-interval-based discovery.
\item We present a mathematical framework to optimize the parameter values of this protocol. This optimization can also be applied to optimize widely-used protocols such as ANT/ANT+ or BLE.
\item We compare the performance of our proposed protocol against time-slotted solutions and show that it can realize significantly lower latency-duty-cycle-products than all known slotted protocols.
\item In real-world measurements, we show that such a protocol can realize the discovery latencies predicted by the theory.
\end{itemize}

\section{The PI-kM Protocol}
\label{sec:PIkMProtocol}
In this section, we first briefly describe the overall scheme of our proposed protocol. Since it contains multiple tunable parameters, we then analyze the properties of this protocol with mathematical methods to derive optimal parameter values. This leads to three variants of the protocols, which are presented in detail.

\subsection{Protocol Overview}
In Section \ref{sec:related_work}, we have described commonly-used asymmetric PI-based protocols like BLE and ANT/ANT+. To become comparable to slotted protocols, we define a symmetric PI-based protocol in which both devices follow exactly the same scheme. Each device repeatedly broadcasts packets with an interval $T_a$ (\textit{advertising interval}) and periodically scans the channel with an interval $T_s$ (\textit{scan interval}) for a duration of $d_s$ time units, as shown in Figure \ref{fig:disco_searchlight_pi_kmp_slot_strategy} c). If a beacon needs to be sent within a period at which the device is scanning, the scanning is interrupted, the beacon is sent and then the device continues scanning without extending $d_s$. We further assume that both devices have been configured with the same values of $T_a$, $T_s$ and $d_s$. However, the theory presented can be extended to one-way discovery (both with different devices roles and different parameter values) easily, thereby making it applicable to existing protocols such as e.g. ANT/ANT+. The duration of one beacon $d_a$ is defined by the number of bytes $N_b$ in each packet and the bitrate $R$. The initial offset between the first advertising packet and the first scan window is distributed randomly between $0$ and $T_s$. Under the common assumption that the device draws the same current for transmission and for reception, the duty-cycle $\eta$ of a device using this protocol is 
\begin{equation}
\label{eq:etaDef}
\eta = \frac{T_a d_s + T_s d_a}{T_a T_s}.
\end{equation}
For the asymmetric case, in which one device only advertises whereas the other one only scans, this Equation describes the joint duty-cycle of both devices. Therefore, the theory presented below remains valid for the asymmetric case, too. For optimizing $T_a$, $T_s$ and $d_s$, the maximum discovery latency $d_m$ needs to be derived. Existing models \cite{kindt:15} rely on recursive schemes which make parameter optimizations difficult. In the following, we derive the worst-case discovery latency considering maxima of order 0 (which means that $T_a < d_s -d_a$) and 1 (which implies that $T_a > d_s - d_a$ and $\gamma < d_s - d_a$), only. For these cases, such a model can be formulated simple enough to derive optimal parametrizations.
All relevant cases for computing the discovery latency are shown in Figure \ref{fig:simpleLatencyComputationModel}. Each case leads to a unique protocol variant.
First, we consider a situation which has only minima of order 0, as shown in Figure \ref{fig:simpleLatencyComputationModel}~a).
\subsection{Case a): The PI-0M-Protocol}
\begin{figure}[htb]
\centering
\includegraphics[width=\linewidth]{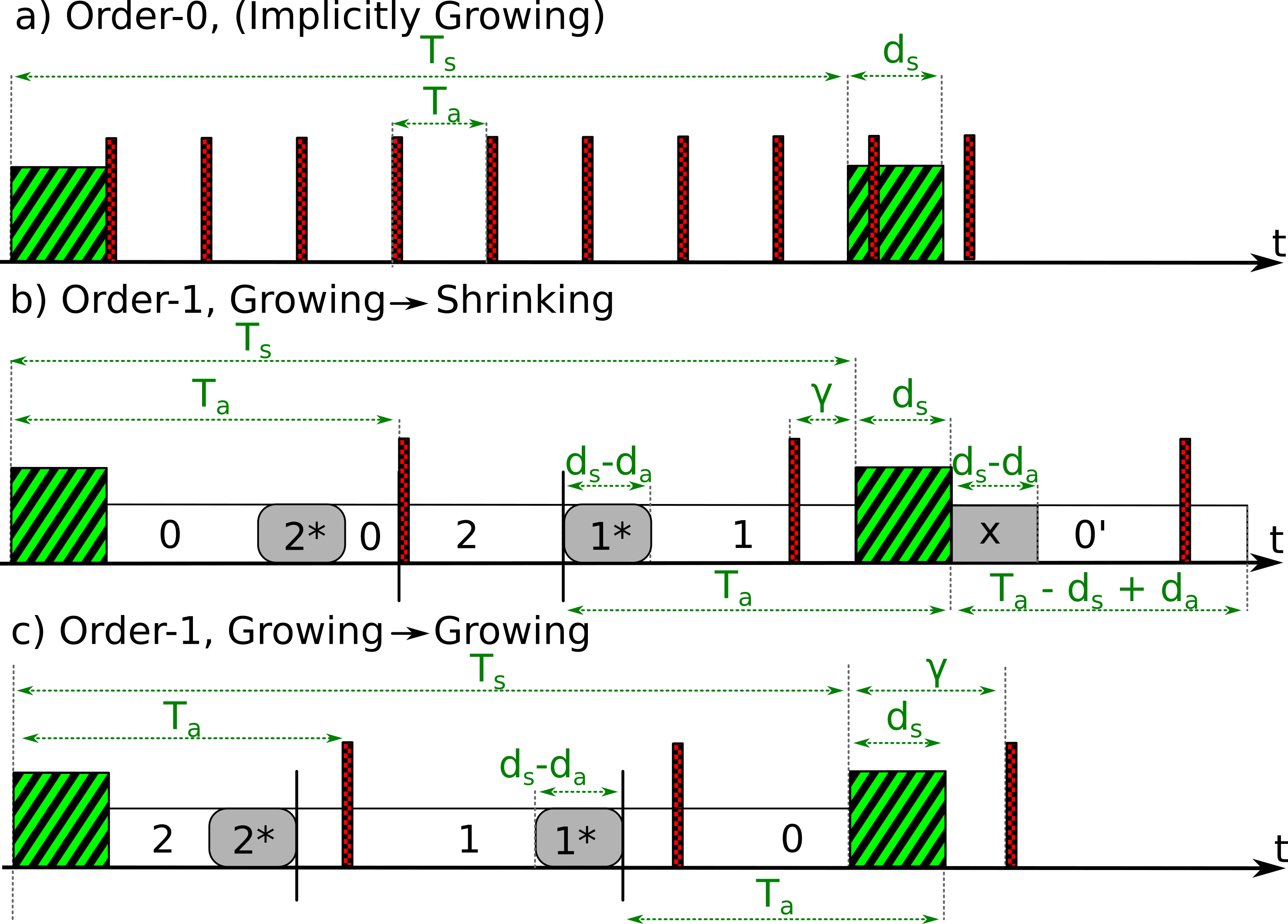}
\caption{Modeling the neighbor discovery latency for a) order-0 and b),c) order-1 problems.}
\label{fig:simpleLatencyComputationModel} 
\end{figure}
Figure \ref{fig:simpleLatencyComputationModel} a) shows the packet flow for $T_a \le d_s$, which is the necessary and sufficient condition to obtain order-0 maxima, only. The ruled boxes depict the scan windows, whereas the hatched vertical bars show the packets. We refer to this protocol as \textit{periodic interval-0M-protocol}, since it has one parameter $M$, as described below.

As can be seen in Figure \ref{fig:simpleLatencyComputationModel} a), $d_m$ is bound to approximately one scan-interval. It can be observed easily that it is\footnote{\label{fn:definition}For our proposed protocols, we define the discovery beginning from the point in time after which both devices have been active for the first time, since this is the latency observed in practical problems. Most other protocols define the latency beginning from the point in time after which the first device has been active for the first time. However, the difference is small.}
\begin{equation}
\label{eq:dMaxPureOrder0}
d_m = \left\lceil \frac{T_s - d_s + d_a}{T_a}\right\rceil T_a + d_a.
\end{equation}
\subsubsection{Choosing $T_a$}
\begin{figure}[htb]
\centering
\includegraphics[width=\linewidth]{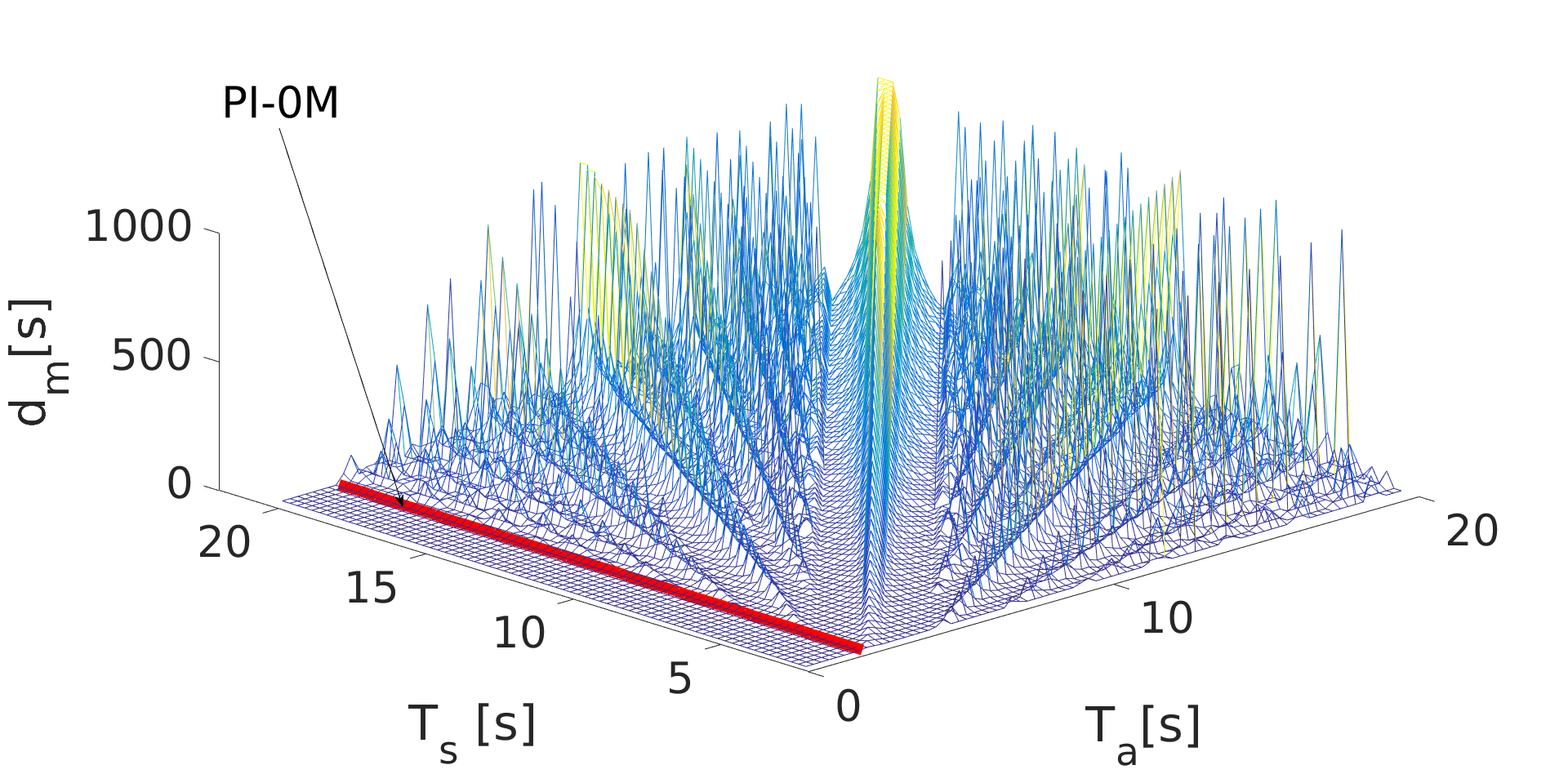}
\caption{Discovery latencies an valuations chosen by Equation \protect\ref{eq:PI0MTaDef} ($\mathbf{d_s = \SI{2.1}{s}}$).}
\label{fig:PI0MTaValues} 
\end{figure}

To minimize the ceiling-function in Equation \ref{eq:dMaxPureOrder0}, it is beneficial to set $T_a$ to its largest possible value. Since we consider order-0 minima, we set
\begin{equation}
\label{eq:PI0MTaDef}
T_a = d_s - d_a,
\end{equation} 
since larger values of $T_a$ would allow for order-1 minima, too. 
Figure  \ref{fig:PI0MTaValues} depicts the maximum discovery latency $d_m$ for sweeping values of $T_a$ and $T_s$ and a fixed value of $d_s$. It also visualizes the parameter values defined by Equation \ref{eq:PI0MTaDef}. As can be seen, they provide low maximum latencies. Smaller values of $T_a$ would lead to similar maximum latencies, but to higher duty-cycles. Therefore, the values chosen by Equation \ref{eq:PI0MTaDef} are optimal for order-0 minima.

\subsubsection{Optimizing $T_s$}
Within certain bounds, the duty-cycle $\eta$ defined by Equation \ref{eq:etaDef} becomes smaller for increasing values of $T_s$. However, if $T_s$ exceeds certain thresholds, the ceiling-function in Equation \ref{eq:dMaxPureOrder0} turns over to its next higher value. Therefore, we attempt to set $T_s$ to its largest possible value under the constraint that the ceiling-function must not increase. Hence, we require:
\begin{equation}
\frac{T_s - (d_s - d_a)}{d_s - d_a} \overset{!}{=} M - \epsilon', M\in \mathbb{N}.
\end{equation}
We assume that $\epsilon'$ is an arbitrary small value.  By solving this, the optimal value of $T_s$ is defined as follows:
\begin{equation}
\label{eq:PI0MTsDef}
T_s = (M+1) (d_s - d_a) - \epsilon.
\end{equation}
In Equation \ref{eq:PI0MTsDef}, we replaced $\epsilon'$ by $\epsilon$. The relation between $\epsilon$ and $\epsilon'$ is not relevant, as long as both values are small. To ease the readability, we assume $\epsilon = 0$ in all subsequent equations and describe its valuations for practical implementations in Section \ref{sec:evaluation}. The maximum latency for parametrizations selected according to Equation \ref{eq:PI0MTaDef} and Equation \ref{eq:PI0MTsDef} is defined as follows:
\begin{equation}
d_m = M (d_s - d_a).
\end{equation}
The desired target duty-cycle defines the value of $d_s$. From Equation \ref{eq:etaDef}, it follows that
\begin{equation}
\label{eq:pi0MdsDef}
d_s(\eta) = d_a + \frac{d_a (M+2)}{\eta (M + 1) - 1}.
\end{equation} 

The only parameter left is $M$. By differentiating the duty cycle-latency-product $\eta \cdot d_m$, it follows that there is a local minimum at
\begin{equation}
M_{opt} = \frac{\sqrt{1-\eta^2} + 1}{\eta} - 1.
\end{equation}

In addition, there are multiple constraints on $M$ which need to be kept. First, $d_s > 0$ and Equation \ref{eq:pi0MdsDef} implies that 
\begin{equation}
\label{eq:pi0MEtaMin}
M > \frac{1}{\eta} - 1 = M_{min}.
\end{equation}

Further, we assume that there is a minimum scan window $d_{s,l}$ the hardware supports. Hence, we require $d_s > d_{s,l}$.
For $\eta < \frac{d_a}{d_{s,l}-d_a}$, this implies
\begin{equation}
M \geq \frac{d_{s,l} (\eta - 1) - d_a (\eta + 1)}{d_a ( \eta + 1) - \eta d_{s,l}}.
\end{equation}
It can be shown that this is always fulfilled if Equation \ref{eq:pi0MEtaMin} is kept. For $\eta > \frac{d_a}{d_{s,l}-d_a}$, it is required that \mbox{$M \leq M_{max}$} with
\begin{equation}
\label{eq:pi0MEtaMax}
M_{max} = \frac{d_{s,l} (\eta - 1) - d_a (\eta + 1)}{d_a ( \eta + 1) - \eta d_{s,l}}.
\end{equation}
Therefore, we set 
\begin{equation}
\label{eq:pi0M_Mdef}
M = \left\{\begin{array}{ccc} \mbox{round}(M_{opt}),& \mbox{if } & M_{opt} <= M_{max}, \\ \lfloor M_{max} \rfloor, & \mbox{else.} &\end{array} \right.
\end{equation}
Finally, it is required that $\lceil M_{min} \rceil \leq \lfloor M_{max} \rfloor$. This translates into the (conservative) requirement 
${M_{max} - M_{min} \geq 1}$. With Equations \ref{eq:pi0MEtaMin} and \ref{eq:pi0MEtaMax}, an upper limit on the supported duty-cycle of this protocol can be derived as follows:
\begin{equation}
\eta_{max} = \frac{d_a + \sqrt{d_a d_{s,l}}}{d_{s,l} - d_a}.
\end{equation}

\subsection{Case b) and c): The $\mathbf{PI-kM^{+}}$-Protocols}

The assumption ${T_a < d_s - d_a}$ is quite restrictive. We therefore allow larger values of $T_a$, as long as ${\gamma < d_s - d_a}$. Such valuations lead to order-1 processes, in which the distance $\Phi[i]$ shrinks or grows with multiples of $\gamma$ until reaching a scan event. Figure \ref{fig:simpleLatencyComputationModel} b) shows the situation in which the offset shrinks in each scan-interval, whereas Figure \ref{fig:simpleLatencyComputationModel} c) depicts the case in which it grows. In this figure, the first advertising and scan-event start at the same point in time. Therefore, the closest neighboring advertising packet of the second scan-event determines which case applies: If the temporally left neighbor is closer than the right one, the process is of shrinking type (case b)). Otherwise, it is growing (case c)). In Figure \ref{fig:simpleLatencyComputationModel}~b) and c), the space in  between the scan-events is subdivided into multiple parts. If an advertising packet starts within one of the colored rounded rectangles, the order-0 process causes a deterministic match after the number of $T_a$ - intervals depicted in each box. These boxes are marked with a *. The other, non-colored rectangles contain the number of $T_a$-intervals until reaching the temporally closest location next to the second scan-event. From this location, the distance is further reduced in multiples of $\gamma$. The value of $\gamma$ is the distance from the second scan event to its closest neighboring advertising packet, as shown in the figure. In case b), the situation is shrinking and hence the distance is reduced from the right side of the scan-event until the advertising packet reaches it. For case c), the distance is reduced in multiples of $\gamma$ from the left side. 

For example, in case b), the transmission of the first advertising packet might start e.g. in box 2. Then, two advertising intervals take place until reaching the field marked with $0'$ in the figure. From there, the temporal distance to the second scan event is reduced in multiples of $\gamma$ until it is reached. Another example is an advertising packet which starts in the box marked with $0$. From there, the distance directly shrinks with multiples of $\gamma$ towards the first scan event. The box marked with \textit{x} can never be reached by any advertising packet.
With these considerations, one can derive the worst-case latency for case b) as follows:
\begin{equation}
\label{eq:dMaxPureOrder1s}
d_m = \left\lceil \frac{T_s - T_a}{T_a} \right\rceil T_a + \left\lceil \frac{T_a - (d_s - d_a)}{\gamma} \right\rceil \cdot \left\lfloor \frac{T_s}{T_a} \right\rfloor T_a + d_a.
\end{equation}
For case c), it is similarly:
\begin{alignat}{2}
\label{eq:dMaxPureOrder1g}
d_m = &\left\lceil \frac{T_s + d_s - d_a}{T_a} - 1 \right\rceil T_a + \nonumber\\
	  & \left\lceil \frac{T_a - d_s + d_a}{\gamma} \right\rceil \cdot \left\lceil \frac{T_s}{T_a} \right\rceil T_a + d_a.
\end{alignat}

\subsubsection{Choosing $T_a$}
\begin{figure}[htb]
\centering
\includegraphics[width=\linewidth]{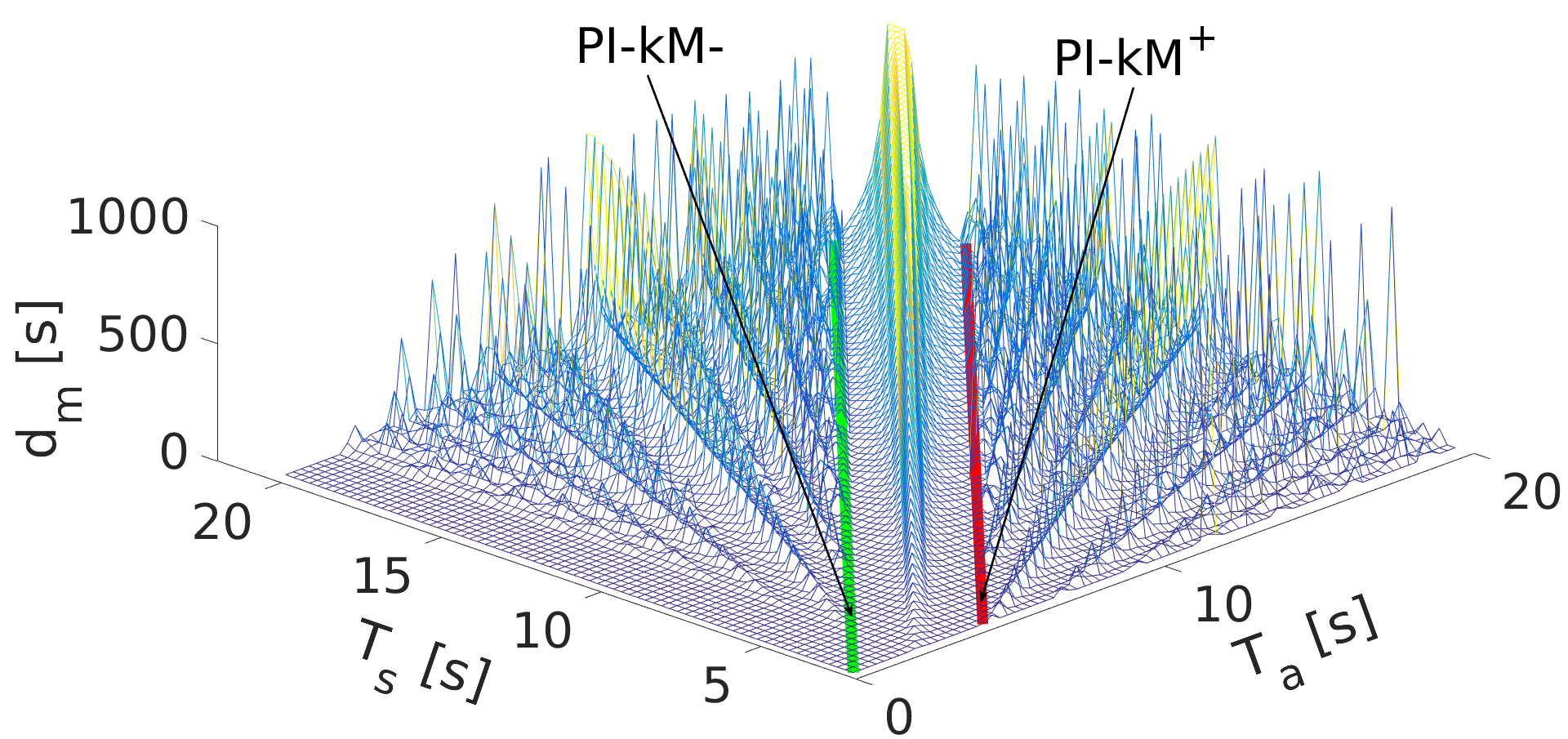}
\caption{Values and associated maximum discovery latencies chosen by Equation \protect\ref{eq:choosingTa}.}
\label{fig:PI-kMPTaValues} 
\end{figure}

When examining Equations \ref{eq:dMaxPureOrder1g} and \ref{eq:dMaxPureOrder1s}, it is clear that the minimum worst-case latencies are achieved for maximum values of  $\gamma$. The maximum value of $\gamma$ is $d_s - d_a$, since otherwise, processes of order 2 and higher would take place. Hence, the minimum worst-case latency is reached for $\gamma = d_s - d_a$.
From Figure \ref{fig:simpleLatencyComputationModel}~b) and c), it can be seen that $\gamma$ is the difference between the second scan event and its closest neighboring advertising packet. It follows that this value can be realized by setting the advertising interval by $d_s - d_a$ time-units shorter or longer than one scan-interval:
\begin{equation}
\label{eq:choosingTa}
T_a = T_s \pm (d_s - d_a).
\end{equation}
Figure \ref{fig:PI-kMPTaValues} shows the worst case discovery latencies for $d_s = \SI{2.1}{s}$, $d_a = \SI{246}{\micro s}$ and sweeping values of $T_s$ and $T_a$. Further, the values of $T_a$ defined by Equation \ref{eq:choosingTa} are highlighted. As can be seen, they all lie within latency minima.  

We refer to the protocol which sets $T_a$ by approximately one packet length longer than $T_s$ as the $PI-kM^{+}$-protocol (because of the $+$-operation in Equation \ref{eq:choosingTa}), and to the protocol which sets $T_a$ slightly smaller than $T_s$ as $PI-kM^{-}$. As can be seen in Figure \ref{fig:PI-kMPTaValues}, these valuations always lie within local latency minima. In the following, we analyze valuations following the $PI-kM^{+}$-scheme in detail. The optimization is more elaborate than for the $PI-0M$-protocol, since there is another parameter $k$, which is used to define additional valuations of $T_a$ for achieving even lower maximum latencies. It is introduced next.

\subsubsection{Optimizing $T_a$}
Equations \ref{eq:dMaxPureOrder1g} and \ref{eq:dMaxPureOrder1s} can be further minimized by reducing $T_a$ without modifying the value of $\gamma$. 
The main idea is that dividing $T_a$ by an integer number $k$ does not change $\gamma$, as can be comprehended by inserting additional packets into the situations shown in Figure \ref{fig:simpleLatencyComputationModel}~b) and c). This measure reduces the maximum distance $\Phi[0]$ which needs to be shrunken in multiples of $\gamma$, and therefore reduces the maximum discovery latency. Instead of the interval length defined by Equation \ref{eq:choosingTa}, we choose $T_a$ as follows.
\begin{equation}
\label{eq:choosingTak}
T_a = \frac{1}{k} (T_s + d_s - d_a).
\end{equation}
\begin{figure}[tb]
\centering
\includegraphics[width=\linewidth]{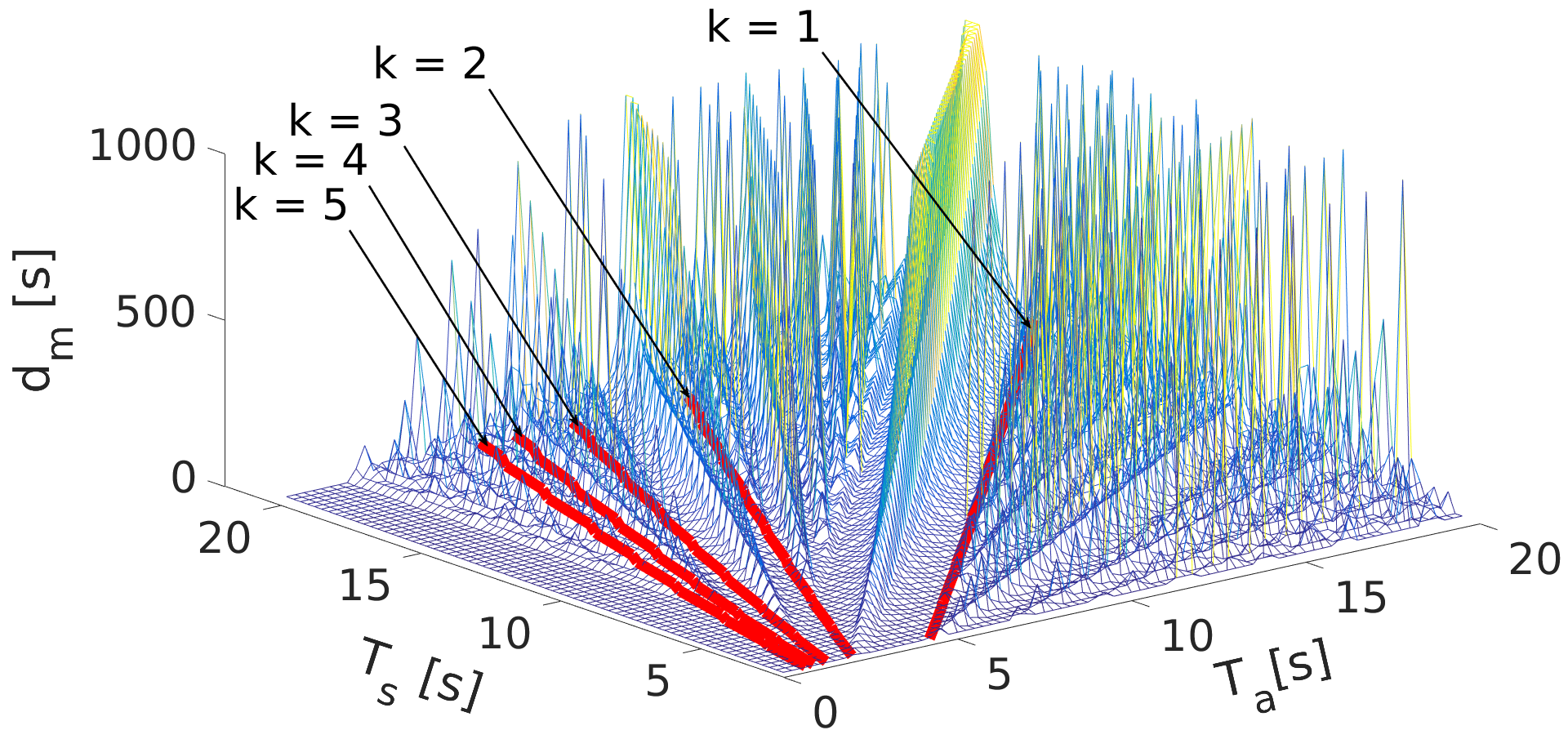}
\caption{Impact of choosing $k$ in Equation \protect\ref{eq:choosingTak}.}
\label{fig:k1to5} 
\end{figure}
Figure \ref{fig:k1to5} shows the values and maximum latencies when choosing $T_a$ according to Equation \ref{eq:choosingTak}. Increasing values of $k$ reduce the maximum discovery latencies, but slightly increase the duty-cycles. Hence, the value of $k$ needs to be set such that an optimum is reached. The lower limit of $k$ is $1$, and the upper limit is defined by requiring $T_a > d_a$, which we never reach in practice. 

Since $T_a$ is a function of $T_s$, the optimization of $T_s$ is the next step towards finding an optimal parametrization. 

\subsubsection{Optimizing $T_s$}

\begin{figure}[htb]
\centering
\includegraphics[width=\linewidth]{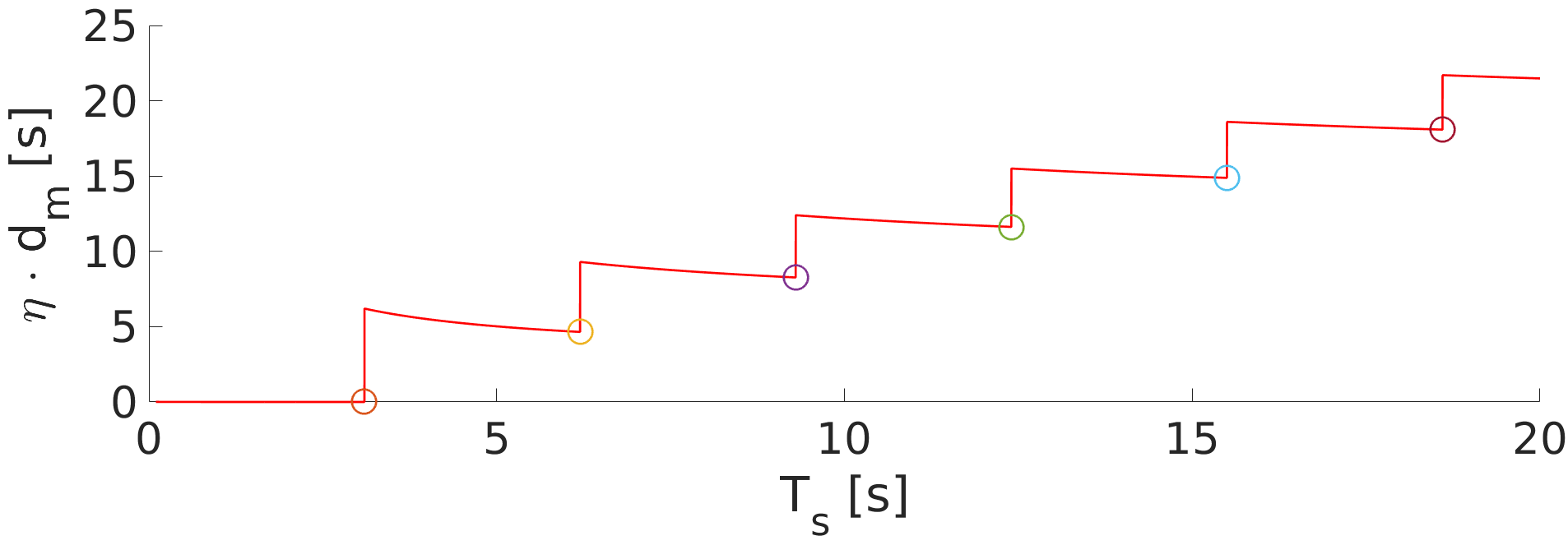}
\caption{The product $\eta \cdot d_m$ for $d_s = \SI{3.1}{s}$, $k=1$ when choosing $T_a$ according to Equation \protect\ref{eq:choosingTak}.}
\label{fig:multipleM} 
\end{figure}

Figure \ref{fig:multipleM} shows $\eta \cdot d_{m}$ for a fixed value of $d_s = \SI{3.1}{s}$ and $k=1$. Pareto-efficient values are the ones depicted by the circles. Since the duty-cycle $\eta$ decreases for increasing values of $T_s$, but $d_{m}$ remains constant as long as the ceiling-functions in Equations \ref{eq:dMaxPureOrder1g} and \ref{eq:dMaxPureOrder1s} do not change their values, we need to find values of $T_s$ which maximize the terms within the ceiling functions, without causing them to turn over.
Hence, we require
\begin{equation}
\label{eq:Ts_optimization_criterion}
\frac{T_s + (d_s - d_a)(1 - k)}{k(d_s - d_a)}  \overset{!}{=} M - \epsilon', M \in \mathbb{N}.
\end{equation}
$\epsilon'$ is a number close to zero. Solving Equation \ref{eq:Ts_optimization_criterion} by $T_s$ leads to:
\begin{equation}
\label{eq:choosingTs}
T_s = (k (M + 1) - 1) (d_s - d_a) - \epsilon
\end{equation}
For convenience reasons, we have introduced a parameter $\epsilon$ instead of $\epsilon'$, as described for the $PI-0M$-protocol.
Equations \ref{eq:choosingTa} and \ref{eq:choosingTs} define the parametrizations of the $PI-kM^{+}$ - protocol. When combining Equations \ref{eq:choosingTak}, \ref{eq:choosingTs}, \ref{eq:dMaxPureOrder1s} and \ref{eq:dMaxPureOrder1g}, its maximum discovery latency can be computed as follows:
\begin{equation}
\label{eq:genericPIkMLatency}
d_{m} = \left\{\begin{array}{l@{\hspace{0.0em}}l@{\hspace{0.1em}}l@{\hspace{0.1em}}}(M - 1) ((M+1)(d_s - d_a) - \epsilon) + d_a,& \mbox{ if } & k = 1,\\(k(M+1)(d_s - d_a) - \epsilon) \cdot & &\\ \cdot \frac{k(M+1)-2}{k} + d_a,& &\mbox{else.} \end{array}\right.
\end{equation}
When assuming $\epsilon=0$\footnote{The error introduced by this is negligible, since $\epsilon$ is small.}, the duty-cycle can be computed as
\begin{equation}
\eta = \frac{d_s - d_a + k d_a + M(d_s + k d_a)}{(d_s - d_a)(M + 1)(k + kM - 1)}.
\end{equation}
Compared to the initial situation, in which there were three real-valued parameters $T_s$, $T_a$ and $d_s$, now there are one real-valued parameter $d_s$ and two integer numbers $k$ and $M$. All valuations lead to local latency minima. In the following, we describe how these values can be further optimized by selecting the best out of these minima. A value of $M=1$ leads to the situation depicted in Figure \ref{fig:simpleLatencyComputationModel} b), in which the offset $\Phi[i]$ shrinks in multiples of $\gamma$. For values of $M > 1$, situation \ref{fig:simpleLatencyComputationModel} c) occurs, in which $\Phi[i]$ grows in multiples of $\gamma$.
Next, we consider $M = 1$.

\subsubsection{Case b): The $PI-k1^{+}$-Protocol (M = 1)}
In the following, we assume $M=1$ and analyze the resulting properties. 
For a given target duty-cycle $\eta$, we can chose $d_s$ as follows:
\begin{equation}
\label{eq:PIk1DsByEta}
d_s(\eta) = \frac{d_a (2k - 1)(2+2\eta)}{2\eta(2k - 1)- 2}.
\end{equation} 
Still, there is one degree of freedom $k$ left. We determine its value by analyzing the constraints on $d_s$. First, $d_s(\eta)$ needs to be larger than zero, and Equation \ref{eq:PIk1DsByEta} implies that
\begin{equation}
\label{eq:PIk1_constraintA1}
k > \frac{1}{2 \eta} + \frac{1}{2}.
\end{equation}
In addition, we assume that the hardware supports a minimum scan window $d_{s,l}$, and require $d_s(\eta) \ge d_{s,l}$.
We can define a limiting value for $k$ as follows.
\begin{equation}
k_{l} = \frac{2 d_{s,l} (1+\eta) - d_a(1+2\eta)}{4\eta d_{s,l} - 2 d_a (1 + 2\eta)}.
\end{equation}
To satisfy $d_s \leq d_{s,l}$, we need to ensure that
\begin{equation}
\label{eq:PIk1_constraintB}
\begin{array}{lcl}
k \le k_l, &\mbox{if}& \eta < \frac{d_a}{2(d_{s,l} - d_a)},\\
k \ge k_l, &\mbox{if}& \eta > \frac{d_a}{2(d_{s,l} - d_a)}.\\
\end{array}
\end{equation}
For $\eta = \frac{d_a}{2(d_{s,l} - d_a)}$, the hardware-constraint is kept if $d_{s,l} \ge \frac{3}{4} d_a$, which is always true in practice. 

By differential computations, it can be shown that the maximum-latency-duty-cycle-product $\eta \cdot d_m$ has a local minimum at $k = k_{opt}$, which is defined as
\begin{equation}
k_{opt} = \frac{1 + \sqrt{(1 - \eta)(1 + 2\eta)}}{2 \eta} + \frac{1}{2}.
\end{equation}
The strategy for choosing k works as follows.
\begin{asparaitem}
\item We set $k \gets \mbox{round}(k_{opt})$, if no constraint is violated.
\item If any of the constraints mentioned above are violated, we set $k$ to the closest possible value to $k_{opt}$ which is allowed by all constraints.
\end{asparaitem}
With this scheme, there are no degrees of freedom left, and a given target-duty-cycle can be realized with optimal parameters. When comparing Equations \ref{eq:PIk1_constraintA1} and \ref{eq:PIk1_constraintB}, one can derive that the maximum duty-cycle which can be realized by the $PI-k1^{+}$-protocol is given by
\begin{equation}
\label{eq:PIk1_etaLimit}
\eta \le \frac{3 d_a + \sqrt{d_a (d_a + 8 d_{s,l})}}{8 (d_{s,l} - d_a)}.
\end{equation}
Next, we examine $M > 1$.

\subsubsection{Case c): The $PI-k2^+$-Protocol ($M > 1$))}
As we will show below, increasing the value of $M$ above 2 is not beneficial in terms of the latency-duty-cycle product achieved. Since we keep $M$ at $2$, we refer to this protocol as the $PI-k2^{+}$-protocol. Again, we can differentiate $\eta \cdot d_m$ by $k$ to find a local minimum $k_{opt}$:
\begin{equation}
k_{opt} = \frac{1}{M+1} + \frac{\sqrt{(1-\eta)(\eta(M+1)+1)} + 1}{\eta(M+1)}.
\end{equation}
Next, we realize that $\frac{d}{dM} (\eta \cdot d_m)$ given $(k=k_{opt})$ is positive for all $M > 1$. Hence, the slope of the duty-cycle-maximum-latency product is always positive and higher values of $M$ lead to higher values of $\eta \cdot d_m$. Therefore, we set $M$ to its minimal value $2$. 
Further, there are multiple constraints which need to be be kept. 
First, $\eta(d_s)$ needs to be positive, and therefore 
\begin{equation}
k > \frac{\eta + 1}{\eta(M+1)} = k_{min}.
\end{equation}
In addition, the protocol should guarantee that no limits of the hardware are exceeded. We again assume a minimum scan duration of $d_{s,l}$ and require $d_s(\eta) \ge d_{s,l}$. This imposes the following upper limit on k:
\begin{equation}
k < \frac{d_{s,l}}{3 \eta d_s - (3 \eta + 1) d_a} + \frac{1}{3} = k_{max}.
\end{equation}
Since k is an integer value, $\lceil(k_{min}) \rceil < \lfloor k_{max} \rfloor$ must be kept. A conservative but analytically solvable form of this inequality is
\begin{equation}
k_{min} + 1 < k_{max} - 1.
\end{equation}
This leads to a general limit of duty-cycles the $PI-k2^{+}$-protocol can realize. Whereas (almost) arbitrarily small duty-cycles are feasible, the upper limit is 
\begin{equation}
\eta < \frac{3 d_a + \sqrt{d_a (d_a + 8 d_{s,l})}}{12(d_{s,l} - d_a)}
\end{equation}
We again set $k$ to $\mbox{round}(k_{opt})$, or as close to it as allowed by the constraints, respectively.

\subsection{One-Way Discovery}
In our proposed protocol, we have modified the scheme used in ANT/ANT+ or BLE\footnote{For the BLE protocol, our scheme needs to be adjusted to account for the random offset that is added to $T_a$.} to obtain a symmetric version of them. The goal was to make them comparable to symmetric slotted protocols. However, e.g. ANT/ANT+ can be seen as a special case of the protocol described at the beginning of this section, by setting its parameters as follows: on the advertising device, $T_s \gets \infty$ and on the scanner, $T_a \gets \infty$. Therefore, the equations presented in this section can be used for selecting optimized parameter values e.g. for ANT/ANT+. The parameters chosen by our theory would then determine the advertising interval for one device and the scan interval and -window for the other one, thereby optimizing the joint energy consumption of both devices. This is highly relevant, since there is currently now known theory for optimizing the parameters of these protocols. Currently, parameters are chosen based on empirical data \cite{AntChannelSearch:09} or "good guesses".
\section{Evaluation}
\label{sec:evaluation}
In this section, we evaluate our proposed technique. Therefore, we first compare the performances of the three protocols described in the previous section among each other. In addition, we compare the theoretically achieved maximum latencies and channel utilizations of our proposed technique against slotted protocols. Further, we demonstrate that our proposed protocol can be implemented on a radio and achieves the predicted performance in real-world measurements.

\subsection{Hardware Parameters}
In this section, we attempt to derive reasonable values for the packet length $d_a$, the slot size $d_{sl}$ and the scan window length $d_{s,l}$. We assume a Nordic nRF51822\cite{nrf51822:14} radio, which supports multiple protocols such as e.g., Bluetooth Low Energy (BLE). It allows for fast switching between sleep- and active modes and between transmission and reception. Therefore, it is well-suited for implementing our proposed protocols.

\subsubsection{Beacon Length}
In typical neighbor discovery applications, it is not only required to detect that another device generates electromagnetic energy on a channel. Typically, one would like to transmit some information about the device during the discovery procedure, e.g., a device address, a device type and some application data. For example, the BLE location beacon Estimote \cite{estimote:16} typically transmits beacons with 46 bytes length (incl. all overheads). On an nRF58122 radio with an over-the-air symbol rate of \SI{1}{MBit/s}, this would translate to a beacon length of $d_a = \SI{368}{\micro s}$. We assume this value for our evaluation.

\subsubsection{Slot Length}
For comparing the performance of slotted protocols against periodic-interval protocols, there is one major challenge: Slotted solutions define a maximum discovery latency in terms of $N_m$ slots, whereas the $PI-kM$-protocols guarantee an upper latency limit in terms of time passed. $N_m$ is independent from the slot length $d_{sl}$, and therefore the maximum latency of slotted protocols is $N_m \cdot d_{sl}$. The duty-cycle of slotted protocols does not depend on the slot length. Clearly, it is beneficial to set the slot-length as short as possible. In contrast, the performance of periodic interval protocols is not proportional to the smallest possible scan window length $d_{s,l}$. $d_{s,l}$ influences the largest supported duty-cycle and sometimes affects the adjustment of $k$ and $M$ due to hardware-constraints, such that the performance can be reduced. Therefore, it should be minimized as well, but its impact is not crucial. 

For slotted protocols, the minimum slot length $d_{sl}$ is limited by multiple factors. First, there is a fundamental limit of $3 \cdot d_a$, which can be derived from Figure \ref{fig:disco_searchlight_pi_kmp_slot_strategy} a) and b):
Every slot contains two beacons with length $d_a$ and a reception phase in between. The radio must listen for at least $d_a$ time-units, since otherwise the packet cannot be received entirely. Next, packet collisions are an issue. From Figure \ref{fig:disco_searchlight_pi_kmp_slot_strategy} a) and b), it can be derived that the collision probability $P_c$ for slotted protocols is 
\begin{equation}
\label{eq:collisonprobabSlot}
P_c = \frac{3 d_a}{3 d_a + d_s}.
\end{equation}
To obtain a reasonable collision rate of e.g. $\SI{10}{\%}$, a slot length of $27 d_a$ is required, which translates to \\ 
$d_{sl} = \SI{9.936}{ms}$ for $d_a = \SI{368}{\micro s}$. Hence $d_{sl} \approx \SI{10}{ms}$ is a reasonable slot length, which we assume throughout the rest of this paper. This value has been assumed in multiple previous studies, e.g. in \cite{dutta:08} and \cite{qiu:16}.

Another limiting factor is clock skew: the clocks of both devices must not drift by more than one slot length per period. For $d_{sl} = \SI{10}{ms}$, one can show that this is not an issue, even if very inaccurate clocks are considered.

\subsubsection{Scan Window Length}
In Section \ref{sec:PIkMProtocol}, we have assumed that there is a lower limit of the scan-window $d_{s,l}$. Whereas all scan windows larger than $d_a$ time-units could be realized by the hardware, clock skew is the limiting factor. In the first part of our evaluation, we assume that $d_{s,l} = d_{sl} = \SI{10}{ms}$ for comparing the latencies achieved against the latencies of slotted protocols. Next, we demonstrate that $d_{s,l} = \SI{10}{ms}$ is a reasonable scan window length for PI-protocols, since it is sufficient to compensate for clock skew on a real-world implementation.

\subsection{Discovery Latencies of PI-Protocols}
In this Section, we evaluate the maximum discovery latencies of our three proposed protocol variants (viz. $PI-0M$, $PI-k1^+$ and $PI-k2^+$) for a given duty-cycle $\eta$. 
We have set the minimum scan window length $d_{s,l}$ to $10 ms$, as described above. The parameter $\epsilon$ is assumed to be $\frac{1}{\SI{32768}{Hz}}$, since this is the smallest step-size a typical crystal that is used as a sleep clock would support.
\begin{figure}[hbt]
\centering
\includegraphics[width=\linewidth]{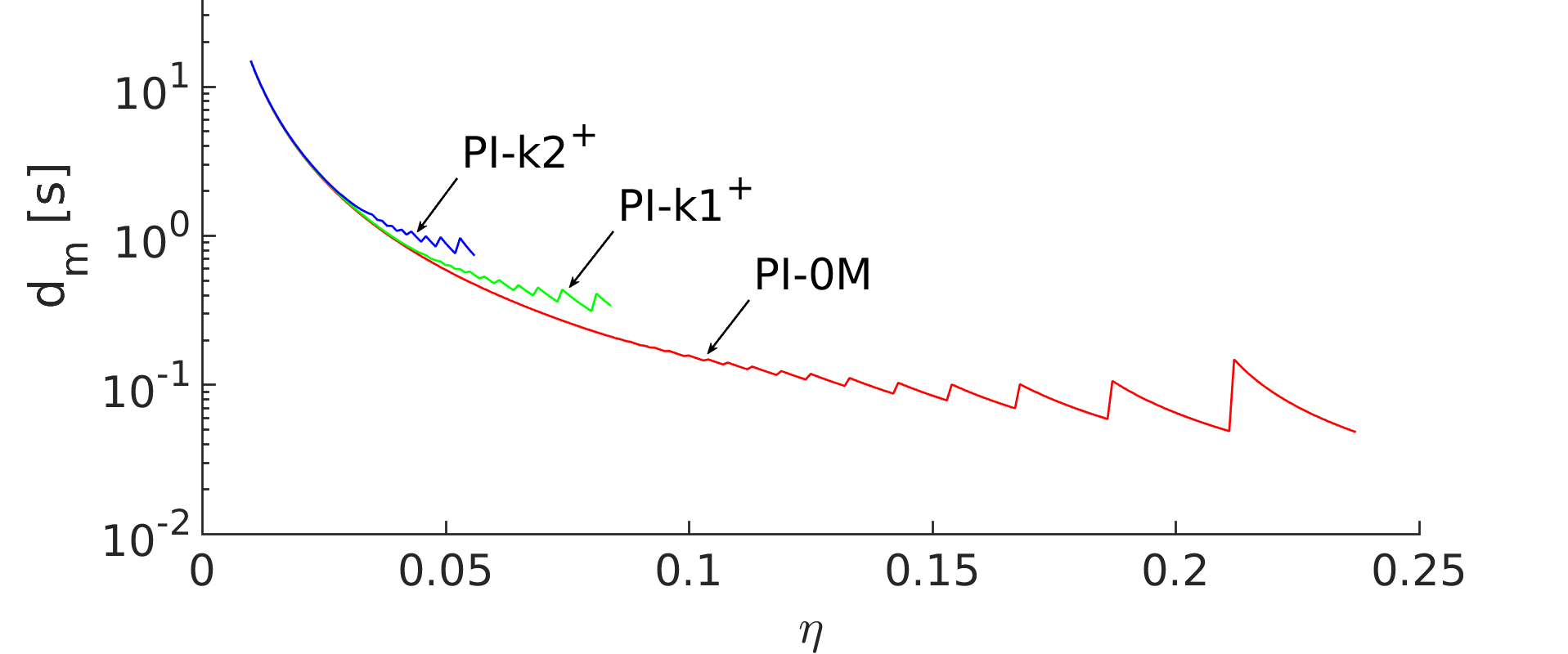}
\caption{Worst Case Latencies $\mathbf{d_m}(\eta)$.}
\label{fig:performance_pi_protocols} 
\end{figure}

Figure \ref{fig:performance_pi_protocols} depicts the maximum discovery latencies $d_m$ of all three protocols for different duty-cycles $\eta$. The latencies have been computed by the Equations in Section \ref{sec:PIkMProtocol}. Starting from $\SI{1}{\percent}$, the evaluation has been carried out for duty-cycles up to $\SI{23.7}{\percent}$, which is the upper limit the protocols can realize for $d_{s,l} = \SI{10}{ms}$. As can be seen, for low duty cycles, the performances of all three protocols are similar. The $PI-0M$-protocol offers the best latencies for larger duty-cycles. For increasing values of $\eta$, the latency function has non-linearities which are caused by adjusting the protocol parameters to meet the constraints described. For the $PI-0M$-protocol, these nonlinearities can become quite significant at larger duty-cycles. Therefore, we propose not to increase the duty-cycle beyond $\eta_{adj}$, which is defined as follows.
\begin{equation}
\label{eq:k0EtaAdjustLimit}
\eta_{adj} = \frac{d_{s,l} - d_a \cdot \sqrt{d_a^2 - 6 d_a d_{s,l} + d_{s,l}^2} + d_a^2 - d_{s,l}^2}{2(d_{s,l} - d_a)^2}.
\end{equation}
The limit from Equation \ref{eq:k0EtaAdjustLimit} ensures that $k$ is not adjusted by more than one, which keeps the performance-degradation negligible.

Since all three protocol variants lead to almost identical latencies for small duty-cycles, each of them can be used in practical applications.

\subsection{Comparison against Slotted Protocols}
In the following, we compare the $PI-kM$ protocols against multiple time-slotted solutions. We have chosen the following protocols for our comparison:
\begin{asparaitem}
\item \textbf{Disco} \cite{dutta:08}, because it is one of the first slotted discovery protocols and the most popular concept that relies on the Chinese Remainder Theorem. For the ease of comparison, we set the two adjacent primes $p_1$ and $p_2$ that Disco requires to $p_1 = p_2$ in the equations for computing the latency. Since such a schedule cannot be realized in practice, DISCO performs slightly worse in real-world implementations.
\item \textbf{U-Connect} \cite{Kandhalu:10}, because it is frequently used as a baseline in many comparisons. \cite{Kandhalu:10} proposes dedicated receiving slots (1 in each period) and transmission slots ($(p + 1)/2$ slots in each hyper-period). This concept allows for extremely short receiving slot lengths of $\SI{250}{\micro s}$, but requires large transmission slot lengths. However, it can only be realized if no payload information is transmitted during the discovery. In addition, it leads to high channel utilizations. We therefore assume a slotting scheme as depicted in Figure \ref{fig:disco_searchlight_pi_kmp_slot_strategy} a), with $d_{sl} = \SI{10}{ms}$.
\item \textbf{Searchlight} \cite{bakht:12}, since, to the best of our knowledge, it is the most efficient slotted symmetric discovery protocol which has been proven to be realizable on real hardware. We consider the striped version of Searchlight, since it offers the lowest discovery latencies. We assume that the overflow $\delta$ is $0$ in all computations, which means that Searchlight performs slightly worse in practice.
\item \textbf{Optimal Diffcodes} \cite{meng:14} provide a theoretical limit which is, to the best of our knowledge, the lowest limit for slotted protocols known so far for most duty-cycles. Despite this limit has not been reached except for a few duty-cycles in practice, we include this theoretical limit into our comparison. The overflow $\delta$ is again assumed to be 0. 
\item \textbf{Lightning} \cite{wei:16}, since it claims to achieve the lowest latency bounds of all known protocols, but has not been evaluated on hardware, yet. We assume the parameters proposed in \cite{wei:16}, i.e. $\beta = \delta = 0.1$.
\item \textbf{G-Nihao} \cite{qiu:16}, since it it constructs similar schedules as the PI-0M+ - protocol. Since G-Nihao provides a parameter to adjust the number of beacons per period, but does not come with any mechanism to find the optimal number of beacons per slot, we assume $\gamma = \frac{n}{m} = 2$, as assumed in the comparison in \cite{qiu:16}. This means that the number of beacons per period $n$ is twice the number of slots per period $m$. It needs to be mentioned that optimized parametrizations of G-Nihao are expected to achieve higher performances. However, the optimal number of beacons per period is not clear.
\end{asparaitem}
Since this study focuses on symmetric asynchronous discovery, we restrict our evaluation on the symmetric variants of the protocols mentioned above.

Under these assumptions, the worst-case discovery latencies for these protocols are as shown in Table \ref{tab:discovery-latency-table}\footref{fn:definition}. 
\begin{table}
\begin{center}
\caption{Worst-Case Discovery Latencies of Slotted Protocols \protect\cite{dutta:08}, \protect\cite{Kandhalu:10}, \protect\cite{bakht:12}, \protect\cite{meng:14}, \protect\cite{wei:16}.}
\label{tab:discovery-latency-table}
{\renewcommand{\arraystretch}{1.3}
\begin{tabular}{cc}
\hline Protocol & $d_m(\eta)$ \\ 
\hline Disco & $\frac{4}{\eta^2} d_{sl}$ \\ 
 U-Connect & $(\sqrt{\frac{1}{2 \eta} + \frac{9}{16 \eta^2}} + \frac{3}{4 \eta})^2 d_{sl}$ \\ 
Searchlight & $\left\lceil \frac{\left\lfloor\frac{1}{\eta}\right\rfloor}{2}\right\rceil d_{sl}$\\
Optimal DiffCodes & $\frac{1}{2 \eta^2} d_{sl}$\\
Lightning & $\frac{n(1+\delta) + (n-1) \delta \beta + 1 + 2 \delta}{\eta - \frac{(1-\delta)\beta + \delta}{2 n (n + 1)} - \frac{\delta + \beta(1-\delta)}{2(n+1)}}$,\\
  & $n \approx \frac{0.0021\sqrt{25600 \eta + 2109} + 0.095}{\eta}$ \\
G-Nihao  & $ \left(\frac{d_{sl} + d_a \gamma}{2 \gamma \eta d_{sl}} + \sqrt{\frac{d_{sl} + d_a \gamma}{2 \gamma \eta d_{sl}}- \frac{d_a}{d_{sl}}}\right)^2 \gamma$ \\
\hline 
\end{tabular} 
}
\end{center}
\end{table}
Among the three proposed protocols $PI-0M$, $PI-k1^+$ and $PI-k2^+$, we always chose the lowest latency for a given duty-cycle\footnote{In our evaluation, the $PI-0M$-protocol always performed best.} and refer to the resulting duty-cycle-latency relation as the $PI-kMOpt$-protocol.

\subsubsection{Discovery Latencies}
\label{sec:discovery_latencies}
\begin{figure}[hbt]
\centering
\includegraphics[width=\linewidth]{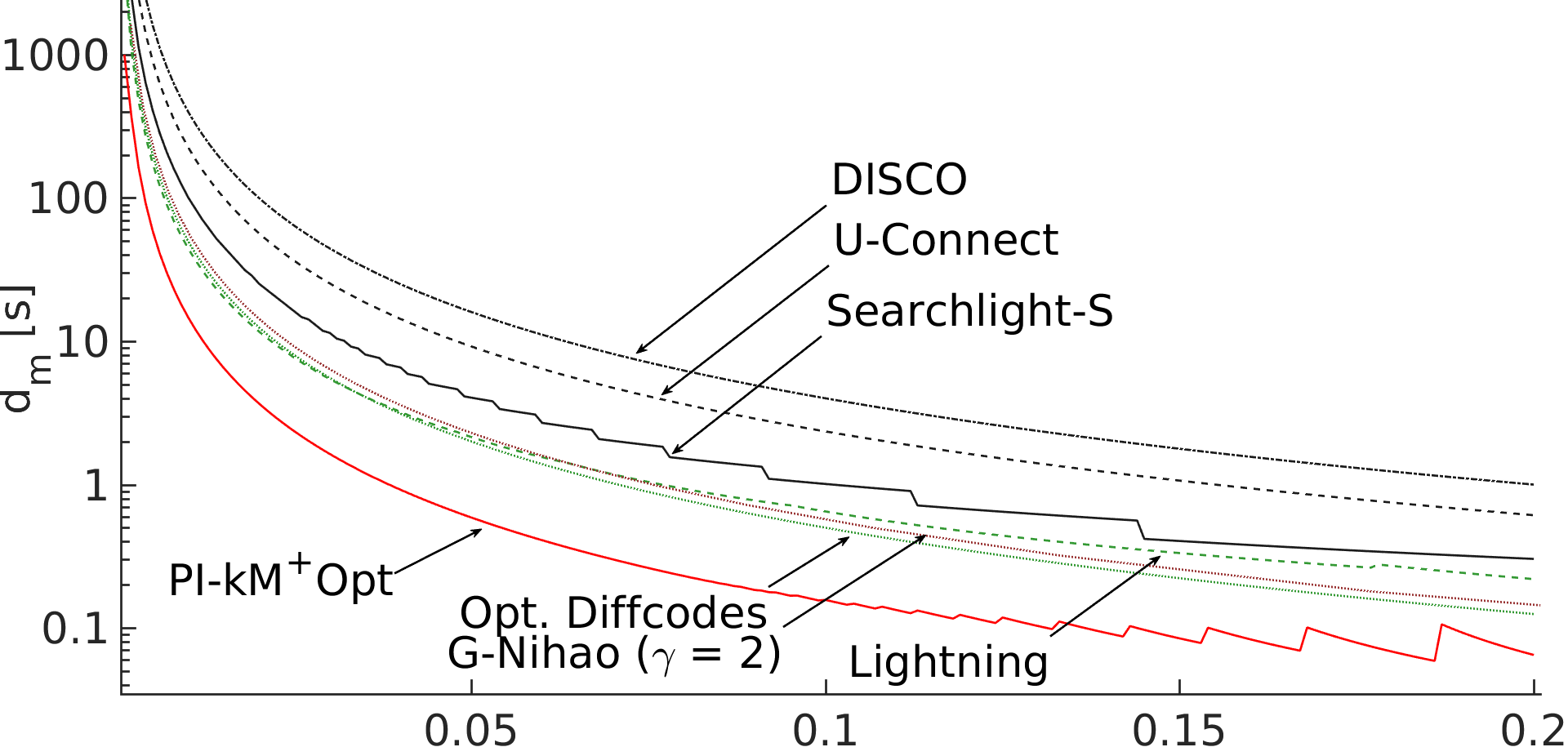}
\caption{Worst-case latencies of slotted protocols against the $PI-kMOpt$-Protocol.}
\label{fig:performance_pi_protocols_vs_slotted} 
\end{figure}

In this section, we compare the discovery-latencies of the $PI-kMOpt$-protocol to the ones achieved by slotted protocols for a relevant range of duty-cycles between $\SI{1}{\percent}$ and $\SI{20}{\percent}$. The results of this comparison are depicted in Figure \ref{fig:performance_pi_protocols_vs_slotted}.
As can be seen, the $PI-kMOpt$-protocol provides significantly shorter discovery-latencies than all slotted protocols for all duty-cycles. For larger duty-cycles, the performance of the $PI-kMOpt$-protocol gets slightly deteriorated due to the adjustments of the parameters for meeting the hardware constraints. 
Table \ref{tab:pi_vs_slotted_results} shows the maximum gains $G_m$ and the mean gains $\overline{G}$ over slotted protocols, defined as $G = \frac{d_{m,slotted}}{d_{m,PI-kMOpt}}$.
For example, for a given duty-cycle, in the worst case, Searchlight would take 10.2 times as long as the $PI-kMOpt$-protocol for discovering a neighbor. 
As already mentioned, G-Nihao constructs similar schedules to the ones constructed by the $PI-0M$-protocol using pseudo-slots. Whereas it has been shown in \cite{qiu:16} that such protocols can guarantee a by factor of 1.65 faster discovery than e.g. Searchlight-S with the same duty-cycle, it has not been studied how such protocols perform in the optimal case, when allowing higher duty-cycles. Our results indicate that the PI-kMOpt-protocol outperforms G-Nihao (with $\gamma = 2$) by up to a factor of $3.1$ at the cost of a higher channel utilization. However, the channel utilization is still within reasonable bounds, as we will show in the next section.
\begin{table}
\caption{Gain in theoretic Worst-Case Discovery Latencies over Slotted protocols. Values marked with a * indicate the theoretic gains with the modifications descried in Section \protect\ref{seq:implementation}}
\label{tab:pi_vs_slotted_results}
\begin{center}
{\renewcommand{\arraystretch}{1.15}
\begin{tabular}{ccccc}
\hline   & $\overline{G}$ & $G_{m}$ & $\overline{G^*}$ & $G_{m}^*$ \\
\hline
Disco & 23.5 & 40.0 & 22.1 & 40.0\\ 
U-Connect & 13.7 & 22.5 & 12.9 & 22.5 \\ 
Searchlight-SR & 6.0 & 10.2 & 5.7 & 10.2 \\ 
Opt. DiffCodes & 2.9 & 5.0 & 2.8 & 5.0 \\ 
Lightning & 3.7 & 4.3 & 3.4 & 4.2 \\ 
G-Nihao & 2.1 & 3.1 & 2.0 & 2.8 \\ 
\hline 
\end{tabular} 
}
\end{center}
\end{table}

\subsubsection{Channel Utilization}
\begin{figure}[h]
\centering
\includegraphics[width=\linewidth]{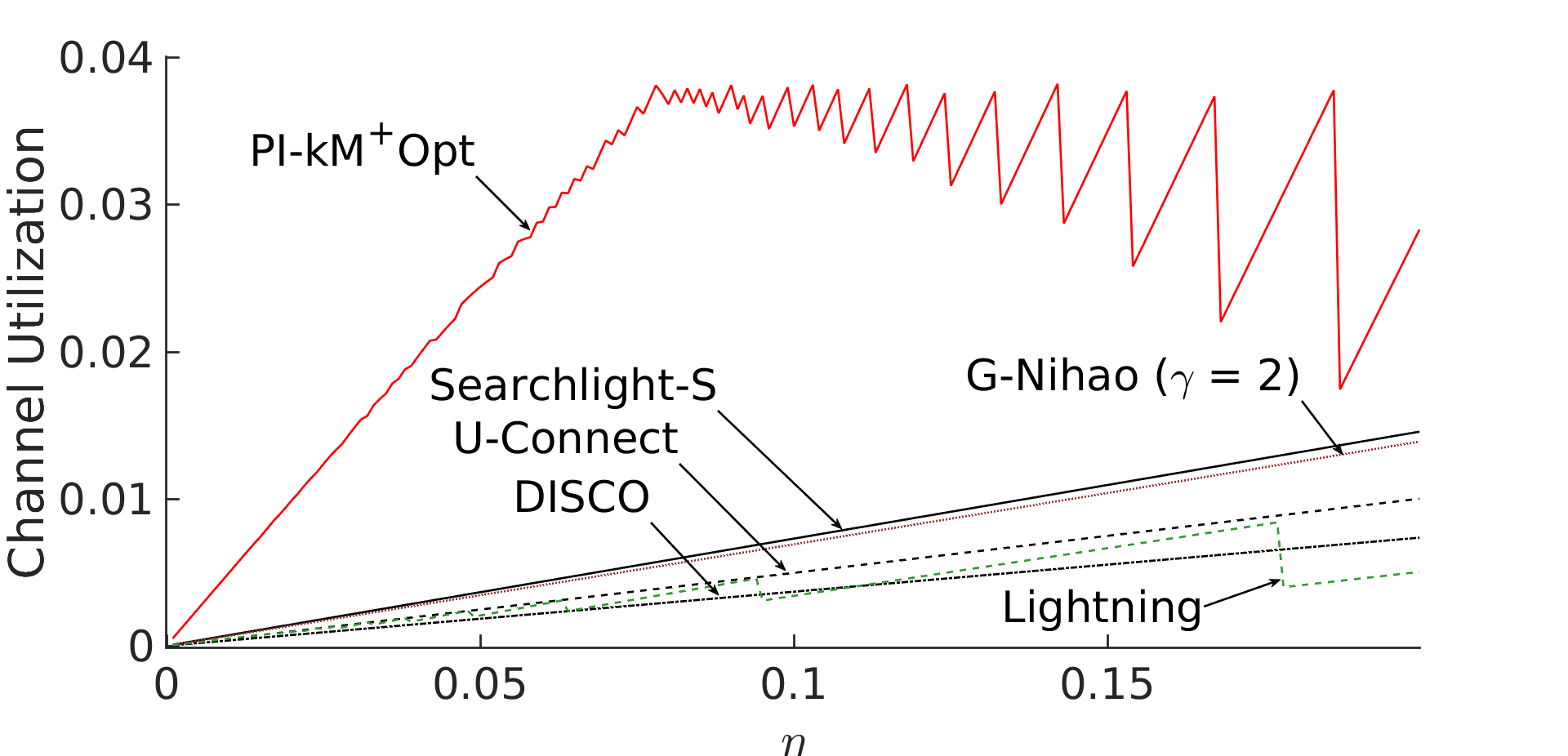}
\caption{Comparison of channel utilizations.}
\label{fig:channel_utilization_pi_protocols_vs_slotted} 
\end{figure}
Unlike in slotted protocols, beaconing and receiving are temporally decoupled in our proposed protocol. Since beacons are short, a large amount of beacons does not increase the duty-cycle significantly, but reduces the maximum discovery latency. However, this could potentially cause large channel utilizations, which would lead to high collision rates and make the protocol impracticable for many situations. To show that the channel utilization is within reasonable bounds, we evaluate and compare it to the utilizations of slotted protocols. The channel utilization is defined by the sum of time-units in which a packet is transmitted on the channel divided by the total time, when considering one device, only. As can be seen in Figure \ref{fig:channel_utilization_pi_protocols_vs_slotted}, compared to the slotted ones, it is indeed increased. However, the total utilization is always below $4 \%$. Between two devices discovering each other, this would lead to a collision rate of up to $\SI{8}{\percent}$. For most duty-cycles, it is significantly lower. It needs to be highlighted that this is an improvement over slotted protocols: The collision rate of all slotted protocols considered is $\SI{10}{\percent}$ for $d_{sl} = \SI{10}{ms}$, and remains constant for all duty-cycles. This advantage is caused by distributing the packets equally over time, whereas for slotted protocols, all beacons are sent temporally compressed within the active slots, only. 

It can be seen that the the pseudo-slotted protocol G-Nihao with $\gamma = 2$ has a similar channel utilization as Searchlight-S, as the authors claim. Nevertheless, it outperforms all other protocols except optimal diffcodes and $PI-kM^+Opt$. 
\subsubsection{Average-Case Behavior}
\begin{figure}[h]
	\centering
	\includegraphics[width=\linewidth]{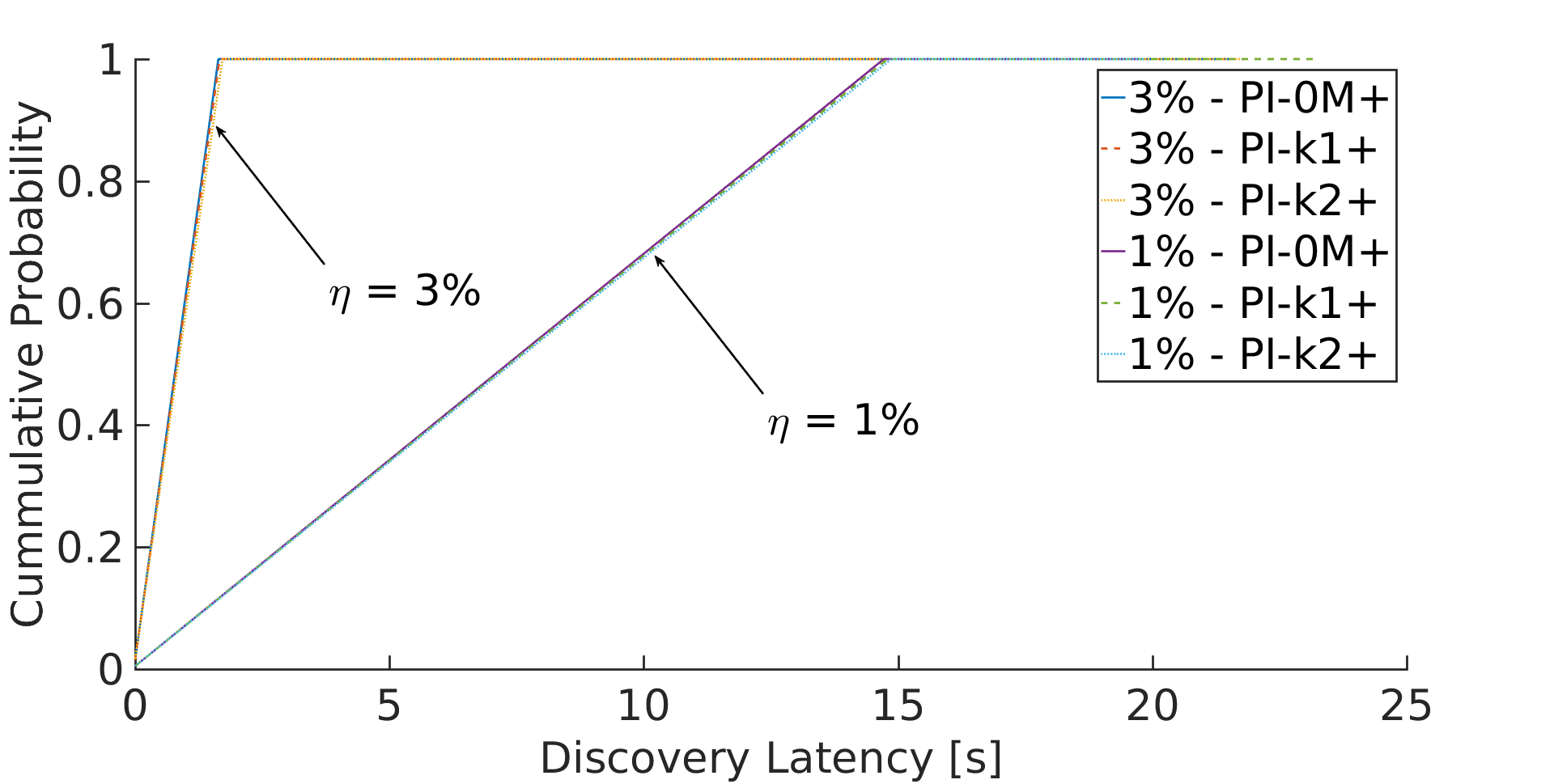}
	\caption{Comulative Distribution Function (CDF) for duty-cycles $\eta = \SI{1}{\percent}$ and $\eta = \SI{3}{\percent}$.}
	\label{fig:PIkM-CDF} 
\end{figure}
Figure \ref{fig:PIkM-CDF} shows that all three protocol variants have nearly identical cumulative distribution functions (CDF). The curves have been generated by computing the latencies for each possible initial temporal offset between the two devices. Since the offsets can be subdivided into multiple fields with constant latencies \cite{kindt:15}, the number of possible offsets is finite and all possibilities can be evaluated in an exhaustive fashion.

As can be observed, the CDFs of the $PI-kM^+$ protocol family are linear functions, which means that their average-case discovery latencies are half of the worst-case latencies.

Compared to the slotted protocols considered in this paper, the average-case performance gains of our proposed protocols are as follows.
The non-randomized version of Searchlight, as well as U-Connect have nearly linear CDF functions \cite{bakht:12}, \cite{kandhalu:13}. Therefore, the performance gains in terms of average latencies are similar to the ones presented in terms of the worst case latencies. The CDF functions of Disco, Lightning and Optimal Difference Codes \cite{dutta:08}, \cite{wei:16}, \cite{meng:14} are concave downwards. Therefore, the performance gains of our proposed protocols regarding the average latencies will exceed the gains regarding the worst-case latencies presented in Section \ref{sec:discovery_latencies}. 

\subsection{Duty-Cycle Granularity}
\begin{figure}[h]
	\centering
	\includegraphics[width=\linewidth]{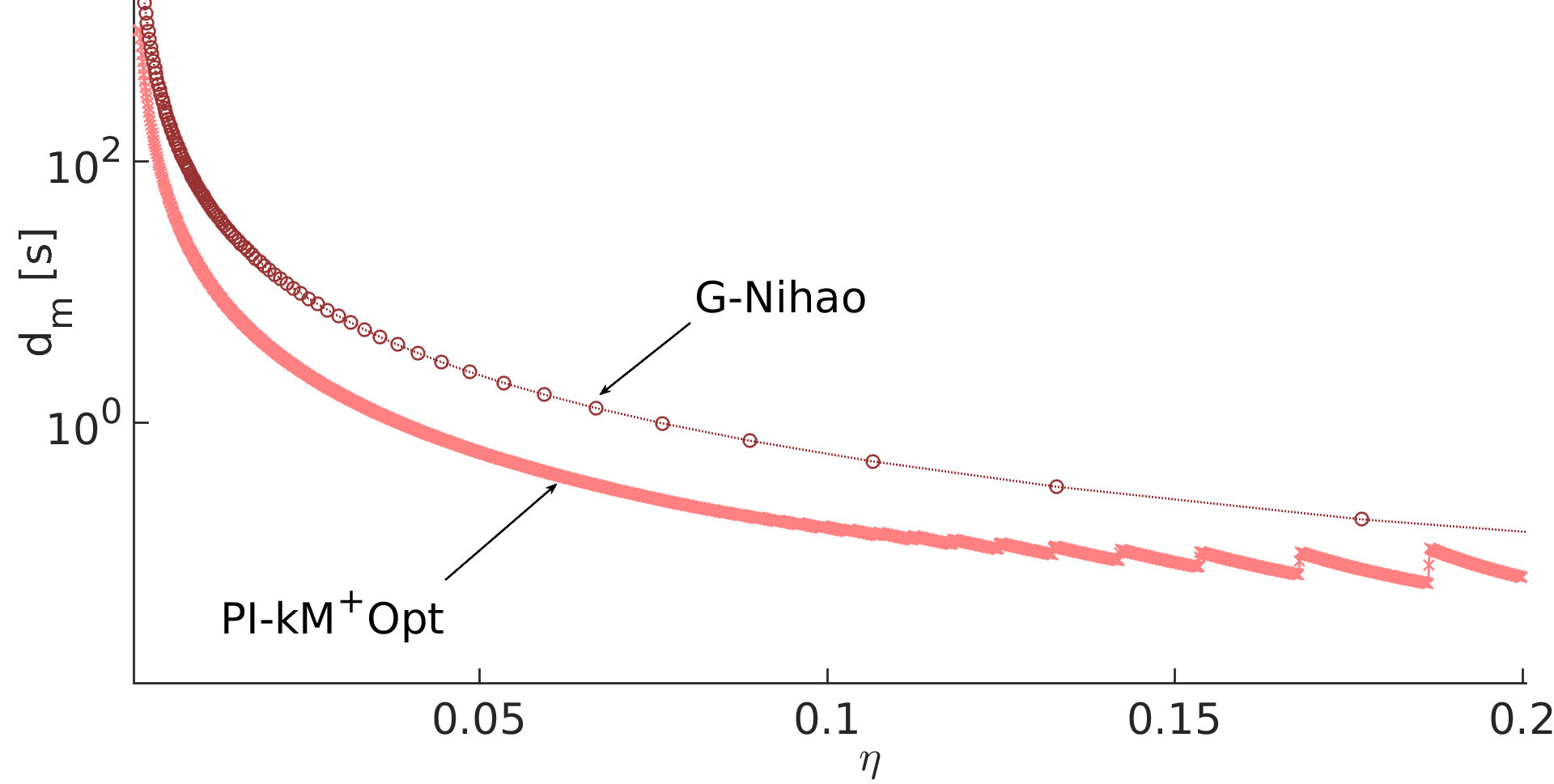}
	\caption{Realizable Duty-Cycles of $PI-kM^+Opt$ and G-Nihao. Every circle depicts a realizable duty-cycle of G-Nihao, whereas every cross depicts a realizable duty-cycle of $PI-kM^+Opt$.}
	\label{fig:dutyCycleGranularity} 
\end{figure}
Besides shorter latencies, slotless protocols have a better duty-cycle granularity than slotted ones. In particular, slotless protocols can realize every possible duty-cycle in practice, and the only limit is the granularity supported by the timers of the radio. To verify this, we have conducted the following experiment. 

A large number of target duty-cycles $\eta_t$ have been defined by sweeping $\eta_t$ between $\SI{0.1}{\percent}$ and $\SI{20}{\percent}$, which is a relevant range in practice (regarding the resulting latencies). We attempted to realize all of these target-duty-cycles both with our proposed protocol, and with one based on a slotted approach.
Due to the already-mentioned similarities, we have evaluated G-Nihao with a slot-length of $d_{sl} = \SI{10}{ms}$ and $\gamma \frac{m}{n}= 2$, as assumed in \cite{qiu:16}. We have computed the value $m$ to realize these duty-cycles. In addition, we have parametrized the $PI-kM^+Opt$-protocol for these duty-cycles, as described in Section \ref{sec:PIkMProtocol}. 

Since G-Nihao is based on a pseudo-slotted theory, $m$ needs to be an integer value. Therefore, we have rounded the resulting value of $m$ to the nearest integer. This limits the duty-cycles that can be realized. Figure \ref{fig:dutyCycleGranularity} shows the results of this experiment. Every realizable duty-cycle of G-Nihao is depicted by a circle, whereas every realizable duty-cycle of $PI-kM^+Opt$ is depicted by a cross. As can be seen, the crosses overlap which each other, since literally every duty-cycle can be realized. In contrast, only a finite number of 
duty-cycles can be realized by G-Nihao. Especially for larger duty-cycles, the granularity decreases significantly. From the literature, it is known that other slotted protocols are even more restrictive. For example, DISCO limits possible duty-cycles to the sum of reciprocals of two prime numbers \cite{dutta:08}. Since slotless protocols do not have this limitation, they allow for continuous on-line duty-cycle controllers, which optimize the duty-cycle continuously during run-time, e.g. given the battery level as an input.

\subsection{Real-World Implementation}
\label{seq:implementation}
\begin{figure}[hbt]
\centering
\includegraphics[width=\linewidth]{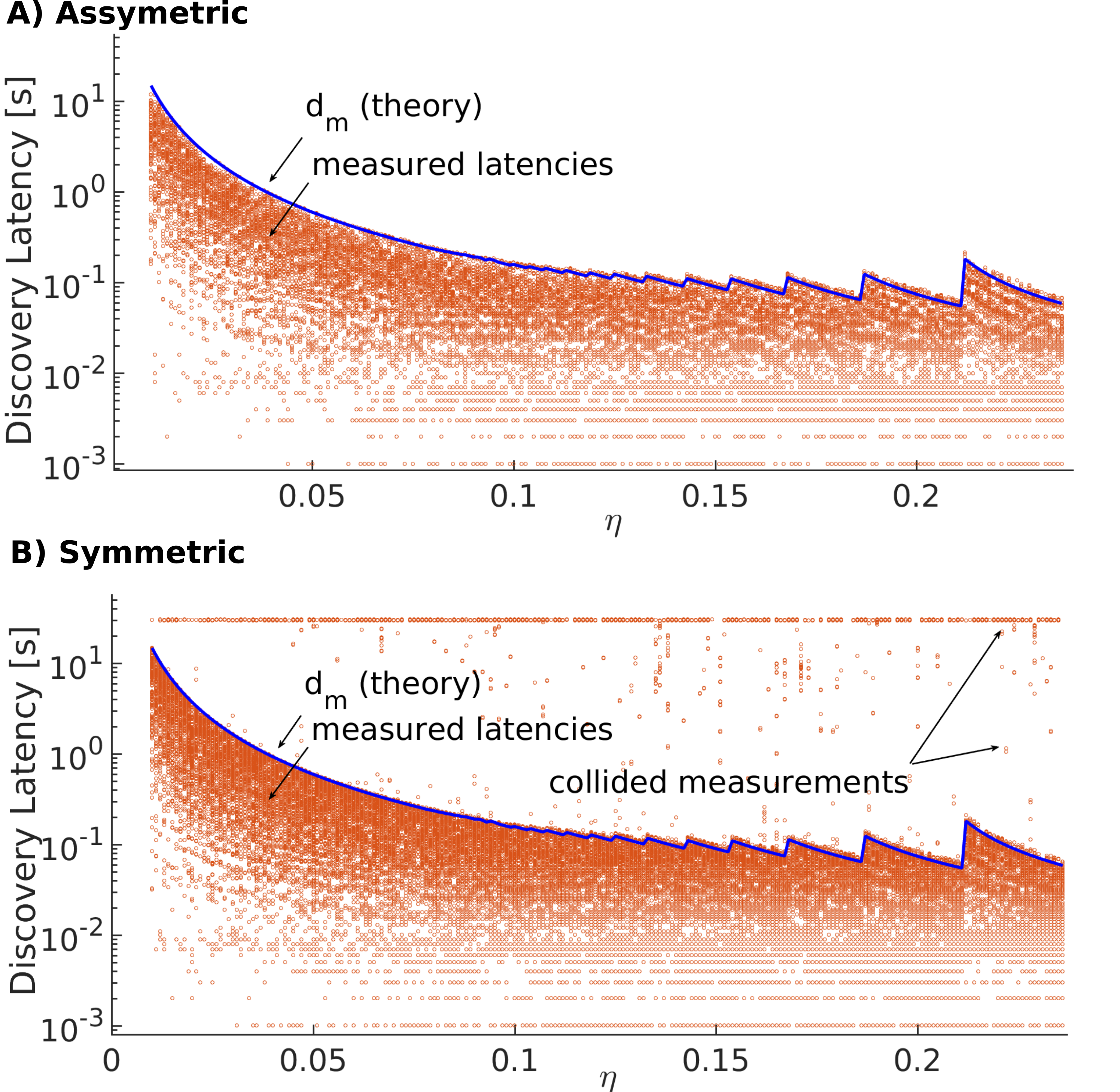}
\caption{Measured against computed values.}
\label{fig:measurement_results} 
\end{figure}
Since the clock of any hardware is subjected to skew, slight adjustments of our proposed protocols are required to compensate for that. These adjustments differ slightly for each of the three protocol variants. We in the following describe them for the most efficient solution, the $PI-0M$-protocol.

First, the ceiling function in Equation \ref{eq:dMaxPureOrder0} might turn over for positive clock skews. Therefore, $\epsilon$ needs to be set such that the largest possible skew $\psi_m$ exceeds this value. For the valuations of $\eta$ we consider, the largest scan-interval that occurs is around $\SI{15}{s}$. When assuming a crystal with an accuracy of e.g. $\SI{20}{ppm}$, one would compute $\epsilon$ as $2\cdot 20 \cdot 10^{-6} \cdot \SI{15}{s} = \SI{600}{\micro s}$, since the clocks of both devices might skew in opposing directions. In addition, the occurrence of higher-order processes needs to be prevented, and therefore ${T_a + \psi_m < d_s - d_a}$. This can be fulfilled by shortening $T_a$ by $\epsilon_{Ta}$, such that $\epsilon_{Ta}$ exceeds $\psi_m$. Therefore, $\epsilon_{Ta} = \epsilon = \SI{600}{\micro s}$. These adjustments slightly reduce the performance of the $PI-kMOpt$-protocol. The gains over slotted protocols marked with * in Table \ref{tab:pi_vs_slotted_results} account for them. As can be seen, the differences to the original gains are small. 

To demonstrate that it is realizable in practice, we have implemented the $PI-kMOpt$-protocol on a radio. Two nRF51822-USB dongles with custom firmwares based on the open-source BLE stack blessed \cite{blessed:15} have been connected to a laptop. In our experiments, each device repeatedly chose a random point in time between $0$ and $T_s$, after which the advertising was started. Once a packet was arrived by a device, it was reported to a laptop which accounted for the points in time this occurred. After either both devices have discovered each other, or a timeout of 30 seconds (which exceeded $d_m$ by more than a factor 2 in the worst case) was reached, the advertising was stopped  and another random offset was chosen for the next round.  For each duty-cycle, the experiment was repeated 100 times. The measured discovery-latencies are shown in Figure \ref{fig:measurement_results} together with the computed upper limits $d_m(\eta)$ (which also accounts for $\epsilon$ and $\epsilon_{Ta}$). Each small point represents the result of one measurement, whereas the solid line depicts the computed limit. In Figure \ref{fig:measurement_results}~a), the asymmetric, one-way case, in which one device advertises and the other one scans, is shown. As can be seen, the measurements confirm our theory. The measured points always lie below the theoretic upper bound. Small deviations are mainly caused by the latency of the USB-connection to the laptop. These results do not only show that the $PI-kMOpt$-protocol can be realized in practice, but also prove that an outstanding performance can be reached by real implementations of it. Figure \ref{fig:measurement_results}~b) shows the measurements of the symmetric, two-way case, in which both devices advertise and scan. As can be seen, the large majority of the measured latencies lies   below the computed curve. However, the predicted maximum latencies are exceeded by a certain fraction of measurements. This behavior is caused by packet collisions. The collision rate has been around $\SI{7.9}{\percent}$, which is close to the expected value. As already mentioned, collisions also occur in slotted protocols. Special channel access strategies (e.g., listen-before-transmit) need to be applied to reduce these collisions \cite{Kandhalu:10}.

\section{Concluding Remarks}
\label{sec:conclusion}
We have introduced a novel discovery protocol, which is based on optimized parametrizations of purely interval-based schemes. It achieves significantly lower discovery latencies than all known slotted protocols. Whereas previous recent studies have shown that breaking away from the assumption of slotted time can increase the performance while maintaining the same channel utilization, we demonstrate that if such protocols are optimized towards performance, much lower latencies than all existing protocols can be guaranteed, while still maintaining acceptable channel utilizations. In addition to its outstanding performance, unlike all previously proposed deterministic protocols, slotless solutions can realize practically every duty-cycle within the range of interest. Given these results, we hope to motivate more researchers to work on slotless protocols in the future.

There is a large potential for future optimizations. For example, starting from Disco \cite{dutta:08}, subsequent slotted protocols performed better than their predecessors because they added additional slots based on a hyper-period. This concept could be seized for interval-based protocols as well. Further, while we focused on valuations with $T_s = T_a + d_s - d_a$, protocols with ${T_s = T_a - (d_s - d_a)}$ need to be evaluated, too.

Besides two-way discovery, the asymmetric one-way case is of high interest, since it is widely used in protocols like BLE and ANT/ANT+. Our proposed optimizations can be applied to these protocols, too. For ANT/ANT+, they can be used directly without any modifications. Therefore, we have presented the first mathematical framework for optimizing its parameters in a systematic fashion. For BLE, small changes have to be applied to account for the random delay and for the discovery on three channels. However, the basic concept remains the same.


\vspace*{\fill}
\pagebreak

\bibliographystyle{IEEEtran}
\bibliography{paper}

\end{document}